\documentclass[12pt]{article}

\usepackage{setspace}
\usepackage[dviwin]{graphicx}
\usepackage{amsfonts}
\usepackage{amssymb}
\usepackage{mathrsfs}
\usepackage{amsmath}
\usepackage{epsfig}
\usepackage{relsize}
\usepackage{graphicx}
\usepackage{mathabx}
\usepackage[textwidth=3cm,backgroundcolor=blue!10]{todonotes}
\usepackage{empheq}
\usepackage{dsfont}
\usepackage{textpos}

\usepackage{lipsum} 
\usepackage{enumitem}
\setlist{nolistsep} 
\usepackage{array}

\newcommand{\bea}{\begin{equation}}
\newcommand{\eea}{\end{equation}}
\newcommand{\bear}{\begin{eqnarray}}
\newcommand{\eear}{\end{eqnarray}}
\newcommand{\bearr}{\begin{eqnarray*}}
\newcommand{\eearr}{\end{eqnarray*}}

\newcommand{\beal}{\begin{align}}
\newcommand{\eeal}{\end{align}}
\newcommand{\beall}{\begin{align*}}
\newcommand{\eeall}{\end{align*}}

\newcommand{\tr}{\mathrm{tr}\,}
\newcommand{\CP}{\mathds{CP}}
\newcommand{\CC}{\mathds{C}}
\newcommand{\dd}{\partial}

\newcommand{\comment}[1]{}

\newcommand{\0}{\textcolor{lightgrey}{0}}

\newcommand{\dP}{\mathbf{dP}}

\newcommand{\can}{\mathcal{K}}
\newcommand{\cst}{\mathrm{J}}

\usepackage{xcolor}

\usepackage{lipsum}

\makeatletter
\def\@seccntformat#1{\@ifundefined{#1@cntformat}%
{\csname the#1\endcsname\quad}
{\csname #1@cntformat\endcsname}
}
\def\section@cntformat{{\normalfont\large\thesection.}\quad}
\def\subsection@cntformat{\textsection\, \thesubsection.\quad}
\def\subsubsection@cntformat{\textsection\textsection\, \thesubsubsection.\quad}
\makeatother

\newsavebox\MBox

\newcommand{\Id}{\mathrm{I}}

\usepackage{hyperref}
\begin{document}

\title{ The K\"ahler metric of a blow-up}
\author{Dmitri Bykov\footnote{Emails:
dbykov@nordita.org, dbykov@mi.ras.ru}  \\ \\
{\small Nordita, Roslagstullsbacken 23, 106 91 Stockholm, Sweden} \\ {\small Steklov
Mathematical Institute of Russ. Acad. Sci., Gubkina str. 8, 119991 Moscow, Russia \;}}
\date{}
\maketitle
\vspace{-0.5cm}
\begin{center}
\line(1,0){450}
\end{center}
\vspace{-0.3cm}
\textbf{Abstract.} After a review of the general properties of holomorphic spheres in complex surfaces we describe the local geometry in the vicinity of a $\CP^1$ embedded with a negative normal bundle. As a by-product, we build (asymptotically locally hyperbolic) K\"ahler-Einstein metrics on the total spaces of the line bundles $\mathcal{O}(-m),\;m\geq 3$ over $\CP^1$. We check that the behavior of the K\"ahler potential is compatible with the Chern-Weil formulas for the Euler characteristic and signature. We also describe two supersymmetric setups where relevant constructions arise.
\vspace{-0.4cm}
\begin{center}
\line(1,0){450}
\end{center}

\begin{textblock}{4}(9,-6.5)
\underline{\small NORDITA-2013-46}
\end{textblock}

\section{Introduction}

This paper grew out of the author's desire to understand certain aspects of the physics of branes placed at singularities of Calabi-Yau varieties. The presentation below, however, will not be in the framework of brane geometry\footnote{Except in \textsection\,\ref{SUSY6D}}, and we will rather concentrate on the intrinsic properties of holomorphic spheres in complex surfaces.

To set up the notations we recall that a complex surface $X$ is an algebraic variety of complex dimension two: $\mathrm{dim}_{\CC}X=2$, and hence of real dimension four. In a local patch we can therefore choose two complex coordinates, $z_1$ and $z_2$. A holomorphic sphere in a complex surface is a copy of $\CP^1$ embedded in $X$ holomorphically. In what follows we will often refer to these as just `spheres'.

In the present paper we will attempt to look at the geometric situation of interest from two points of view: the one of algebraic geometry and the one of differential geometry. We will make an important assumption about the symmetry group of $X$, namely, from the differential geometric point of view, we will require that the metric on $X$ has $U(2)$ as the group of isometries. On the algebraic side this translates into the statement that the automorphism group of $X$ includes $GL(2,\CC)$. It is a restrictive requirement, and it will allow to reduce all partial differential equations to ordinary ones and therefore to simplify the problem tremendously. Due to the $U(2)$ invariance it will be convenient to introduce the combination
\bea\label{x}
x:=|z_1|^2+|z_2|^2
\eea
of the local coordinates $(z_1, z_2)$.

The main result of the present paper is that in the $U(2)$-invariant case the K\"ahler potential in the vicinity of a $\CP^1$ embedded with normal bundle $\mathcal{O}(-m)$ looks as follows:
\bea\label{Kahlerexpansion}
K=a \log{x}+b\, x^m+\ldots \quad \textrm{for}\;\;x\to 0\,,
\eea
where $a, b>0$. In this formula we imagine that the $\CP^1$ is glued in at the origin, i.e. at $x=0$. Intuitively, this is a very clear formula. Indeed, if we drop all corrections to the leading $\log\,x$ term, then we obtain a K\"ahler potential on $\CP^1$, which, in particular, means that it would lead to a  degenerate metric on the surface $X$. Hence, the role of the correction term $x^m$ is that it lifts the degeneracy of the metric in the directions `normal' to the glued-in copy of $\CP^1$.

The motivation for conjecturing (\ref{Kahlerexpansion}) comes from two main examples, in which we consider holomorphic spheres in the following surfaces:
\bear
&& m=1: \quad \textrm{del Pezzo surface of rank 1}\\
&& m=2:\quad \textrm{Eguchi-Hanson space, i.e. resolution of}\;\; xy=z^2\,, \;\textrm{or}\;\CC^2/\mathbb{Z}_2
\eear
There is a tight connection between these two surfaces --- both of them can be thought of as `blow-ups' (hence the title of the paper). Indeed, $\dP_1$ is the blow-up of $\CP^2$ in a single \emph{smooth} point, whereas the resolution of the singularity of $xy=z^2$ at $x=y=z=0$ is a blow-up of this surface at the \emph{singular} point --- the origin. As we have just announced, despite this superficial similarity, the normal bundles to the corresponding $\CP^1$'s which arise as the exceptional divisors of the blow-ups in the two cases are different.

The vicinity of a $\CP^1$ embedded with normal bundle $\mathcal{O}(-m)$ looks like the total space of this line bundle --- we call this space $Y_m$. It is tempting to try to build a metric on $Y_m$, which by definition should exhibit the characteristic behavior (\ref{Kahlerexpansion}) at $x\to 0$. We find that for $m\geq 3$ it is possible to build a K\"ahler-Einstein metric on $Y_m$ rather explicitly. It turns out that it is the Lobachevsky space (i.e. a unit ball in $\CC^2$) with a `glued in' copy of $\CP^1$. Near the boundary of the ball the metric becomes the one of constant negative curvature, i.e. the Lobachevsky metric.

If one analyzes solely the vicinity of the blow-up, a priori there is no restriction on $m$, apart from positivity: $m>0$. However, using Chern-Weil theory, we show that for generic $m$ the Euler characteristic and/or signature of the corresponding surface turns out to be non-integer, which cannot be the case for a smooth manifold. Restricting only to integer topological data leads to the quantization of $m$ and to the condition $m\geq 2$. For $m\geq 2$ it also requires that we identify the points $(z_1, z_2)\sim e^{{2\pi i \over m}}\,(z_1, z_2)$, which in particular implies that the boundary has to be the lens space $S^3/\mathbb{Z}_m$ rather than simply a sphere $S^3$.\vspace{0.3cm}

The paper has the following structure:\newline
\begin{itemize}[itemsep=1pt]
\item \textsection\,\ref{deformation}. We review the connection between deformation properties of holomorphic spheres and their normal bundles.
  \item Section \ref{selfintsec}. We discuss the relation between the self-intersection number $\mathcal{C}.\mathcal{C}$ of a sphere $\mathcal{C}=\CP^1\subset X$ and its normal bundle.
  \item Section \ref{surfaces}. We present the two main examples --- two surfaces --- that motivate the subsequent discussion of the K\"ahler potential in the vicinity of a holomorphic sphere $\CP^1 \subset X$.
   \item \textsection \;\ref{dP1sec}. We introduce our first example --- $\dP_1$ --- the del Pezzo surface of rank 1.
  \item \textsection \;\ref{SUSYKahler}. In a supersymmetric setup we introduce the K\"ahler quotient construction.
  \item \textsection \;\ref{Kahlermath}. We describe the K\"ahler quotient directly, from a mathematical standpoint.
  \item \textsection \;\ref{dP1metr}. Using the K\"ahler quotient, we build a metric on $\dP_1$.
  \item \textsection \;\ref{A1sec}. Our second example is a surface that arises after the resolution of an $\mathds{A}_1$-singularity. We discuss the ideas that underlie the resolution of $\mathds{A}_n$ singularities and provide two examples:
  \begin{itemize}
  \item \textsection \;\ref{a1sing}: $\mathds{A}_1$ (which we use in what follows). We discuss how the self-intersection number $\mathcal{C}.\mathcal{C}$, which was previously introduced geometrically as the degree of the normal bundle to $\mathcal{C}$, can be computed purely by algebraic means.
  \item \textsection \;\ref{a2sing}: $\mathds{A}_2$ (which is of a more illustrative nature): generalizes the previous discussion of $\mathds{A}_1$.
  \end{itemize}
  \item \textsection \;\ref{EH1}. The surface obtained by resolving the $\mathds{A}_1$ singularity may be seen as the total space of the canonical bundle $\mathcal{O}(-2)$ over $\CP^1$. Here we review the Eguchi-Hanson metric --- the Ricci-flat metric on the total space of $\mathcal{O}(-2)$.
  \item \textsection \;\ref{EH2}. Using the hyper-K\"ahler property of the Eguchi-Hanson space, we rewrite this metric in the form of a hyper-K\"ahler quotient.
  \item \textsection \;\ref{SUSY6D}. We review how an identical quotient arises in a supersymmetric $\mathcal{N}=(1, 0)$ theory in six dimensions from field-theoretic considerations.
  \item Section \;\ref{topology}. We review the main topological characteristics of 4-manifolds.
  \item \textsection \;\ref{ChernWeil}. We introduce the Chern-Weil formulas that allow to calculate the topological numbers from the K\"ahler potential.
  \item Section \;\ref{KEmetr}. We show that there exist K\"ahler-Einstein metrics on the total spaces of $\mathcal{O}(-m)$ bundles over $\CP^1$ for $m\geq 3$, of negative curvature. The latter metrics are built explicitly. The topology of the underlying surfaces is studied.
  \item Appendix \;\ref{Kahmomrel}. We derive a relation between the K\"ahler potential and the moment map, that is used in Section \ref{SUSYKahler}.
  \item Appendix \;\ref{appKahquot}. We derive the K\"ahler potential on a K\"ahler quotient variety in terms of the original K\"ahler potential by an explicit calculation.
  \item Appendix \;\ref{ChernSimons}. We derive the `Chern-Simons'-type boundary correction to the Chern-Weil formulas.
  \item Appendix \;\ref{metric}. We write out explicitly the expression for the line element arising from a $U(2)$-invariant K\"ahler potential, that depends only on $x$.
\end{itemize}

\subsection{Spheres with a positive normal bundle: deformation}\label{deformation}

\begin{figure}
\centering
\includegraphics[width=0.8\textwidth]{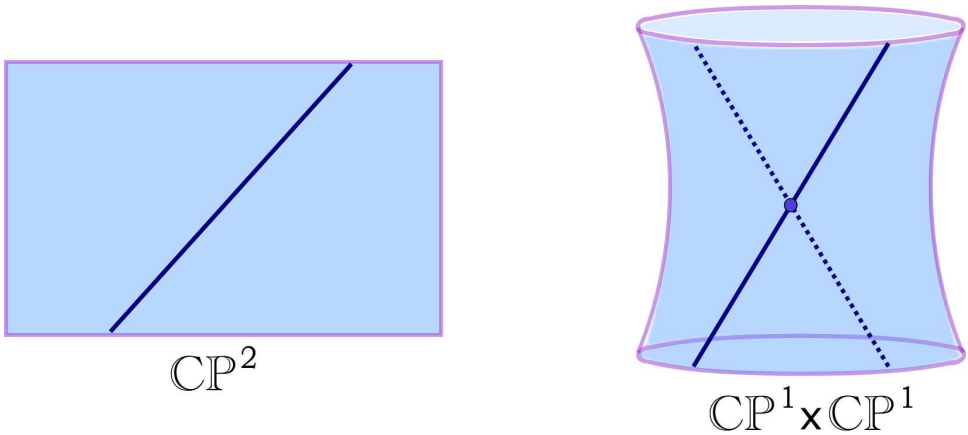}
\caption{Lines embedded with nonnegative normal bundle: \newline  (1) $\CP^1 \subset \CP^2$\;\;(2) $\CP^1 \subset \CP^1 \times \CP^1$.}
\label{linesfig}
\end{figure}

We are used to the fact that there are infinitely many lines in a plane. Moreover, all of them may be obtained from a single line by deformation, i.e. a motion. $\CP^2=\CC^2 \bigcup \CP^1$ is a compactification of the plane. In such a compactification a line in $\CC^2$ is promoted to a holomorphic sphere $\CP^1$. There are situations when the question about the number of $\CP^1$'s in a given algebraic variety is not void of meaning. Clearly, this can only be the case if a given sphere cannot be continuously moved around. A sufficient condition for this is that its normal bundle is negative, meaning that it is isomorphic to $\mathcal{O}(-m)$ for $m>0$. Indeed, for a sphere to have a continuous deformation its normal bundle should have holomorphic sections. This is only the case for a positive (or a trivial) normal bundle. Let us elaborate on the above requirement. Suppose for simplicity that $X$ is given by a single homogeneous polynomial equation
\bea
F(Y_1:Y_2:Y_3:Y_4)=0\;\subset \CP^3
\eea
The statement that there is a sphere in $X$ means that there is a map from $\CP^1$ with homogeneous coordinates $(u_1:u_2)$ to $\CP^3$ with coordinates $(Y_1:Y_2:Y_3:Y_4)$, such that $Y_i$ are homogeneous functions of $u_a$, and the polynomial $F$ is zero on the image of this map. If the line can be deformed, i.e. there is a continuous family of such lines, then the functions $Y_i(u_a)$ depend on some extra parameter $t$, which labels the lines inside the family. If $t=0$ corresponds to the original line $L\subset X$, ${\dd \over \dd t}\big|_{t=0}$ is the normal vector field to $L$ in $X$:
\bea
{\dd \over \dd t}|_{\,t=0}={\dd Y_i \over \dd t}|_{\,t=0}\,{\dd \over \dd Y_i}|_{\,L}
\eea
Clearly, we assume that $\dd Y_i \over \dd t$ is not identically proportional to the tangent vector field(s) $\dd Y_i \over \dd u_a$ (which means it is a genuine normal vector field). At the points (if any) where ${\dd \over \dd t} \,\propto\, {\dd \over \dd u_a}$ the normal vector field effectively has a zero.

\vspace{0.3cm}
\emph{Examples with nonnegative normal bundles} (see Fig. \ref{linesfig}):

\vspace{0.3cm}
(1) $\CP^1$ inside $\CP^2$. Let $(z_1:z_2:z_3)$ be the homogeneous coordinates on $\CP^2$. Then we assume that  $L=\CP^1$ is embedded by means of the equation $z_3=0$. At $L$ the holomorphic tangent bundle to $\CP^2$ splits as $\mathrm{T}\CP^2\big|_L=\mathrm{T}L\oplus \mathrm{N}L$. Taking the determinant, we obtain $\mathcal{O}(3)=\mathcal{O}(2)\otimes \mathrm{N}L$. This imples $\mathrm{N}L=\mathcal{O}(1)$.

\vspace{0.3cm}
(2) $\CP^1$ inside $\CP^1 \times \CP^1$, embedded as $z\in \{L=\CP^1\}\to (z, p)\in \CP^1 \times \CP^1$, where $p$ is a  (chosen) point on the second $\CP^1$. Clearly, the fiber of the normal bundle to $L$ is isomorphic to the tangent space to the second $\CP^1$ at the point $p$, which is the same for any $z\in L$. Hence $\mathrm{N}L=L\times \CC$ is a trivial bundle.

\section{The normal bundle and self-intersection}\label{selfintsec}

The goal of this paragraph is to introduce a useful relation between the self-intersection number $\mathcal{C}.\mathcal{C}$ of a curve $\mathcal{C}\subset X$ and the normal bundle to the curve, i.e. the topological characteristic of how it is embedded in $X$.

Let us consider the simplest type of surface singularity, namely the $A_n$ family, which is given by the following equation:
\bea\label{A-n}
xy=z^{n+1}
\eea
First of all, we focus on the case $n=1$, and we wish to show that $xy=z^2$ describes the singularity of an affine cone over $\CP^1$ with respect to its anticanonical embedding. We choose the homogeneous coordinates $(z_1: z_2)$ on $\CP^1$. Since the anticanonical line bundle over $\CP^1$ is $\mathcal{O}(2)$, we can choose a linear basis in the space of its sections (the so-called `linear system of divisors') as $x=z_1^2, y=z_2^2, z=z_1 z_2$. It is obvious that these variables satisfy (\ref{A-n}) with $n=1$. So far, since we are viewing $(x,y,z)$ as projective coordinates, what we have constructed is simply the Veronese embedding of $\CP^1$ into $\CP^2$. Building an affine cone over this embedding corresponds to forgetting the projective identification of $(x, y, z)$, i.e. regarding them as affine variables --- then we obtain the equation of an $A_1$ singularity in $\CC^3$.

In general, it is known that the resolution pattern for an $A_n$-singularity produces several spheres, intersecting at points according to the Dynkin diagram of $A_n$, i.e. forming a chain. We discuss it in more detail on the example of an $A_2$ singularity in \textsection \,\ref{a2sing}.

Singularities of the type (\ref{A-n}) are particular examples of the so called du Val singularities. The characteristic property of du Val singularities is that blowing them up does not affect the canonical class of the variety \cite{Shafarevich}. The algebraic reason for this is that the copies of $\CP^1$ that we glue in during the blow-up process are the so-called $(-2)$-curves, which means that they have self-intersection number $-2$. In fact, in algebraic geometry there is the following important formula that relates the genus of a smooth curve $\mathcal{C}\subset X$ in a surface $X$ to the intersection numbers of the corresponding divisor, which we will also call $\mathcal{C}$, and the canonical class $\mathcal{K}$ of the surface\footnote{For two complex curves $\mathcal{L}_1, \mathcal{L}_2 \subset X$ we will denote by $\mathcal{L}_1.\mathcal{L}_2$ their intersection number.}:
\bea\label{genus}
g_{\mathcal{C}}=\frac{\mathcal{C}.(\mathcal{C}+\mathcal{K})}{2}+1
\eea
In particular, we see that if $X$ is Calabi-Yau, i.e. when $\can=0$, curves with self-intersection $-2$ have genus zero, hence they are spheres ($g_{\mathcal{C}}=0$).

A del Pezzo surface, by definition, is a surface with ample anticanonical class, which is usually expressed as $-\can>0$. In the opposite situation, a surface of general type has ample canonical class, i.e. $\can>0$. From the so-called Nakai-Moishezon criterion of ampleness it follows that an ample line bundle (or divisor) $\mathcal{L}$ has a positive intersection with any curve $\mathcal{C}$ in $X$, i.e. $\mathcal{C}.\mathcal{L}>0$. Hence for the del Pezzo surface we obtain $\mathcal{C}.\can<0$ and for a surface of general type we get $\mathcal{C}.\can>0$. It is now easy to relate these definitions to the geometric properties of the surface, such as its curvature. Indeed, it is well-known that if $R_{i\bar{j}}$ is the Ricci tensor, the 2-form of type (1,1) ${i\over 2\pi} R_{i\bar{j}} dz^i\wedge d\bar{z}^j$ is closed and belongs to the (integral) cohomology class of $c_1(X)=-c_1(\can)$. If $\omega_\mathcal{C}$ is the 2-form Poincare dual to $\mathcal{C}$, the intersection $\mathcal{C}.\can$ may be interpreted in differential geometry as the integral\footnote{For a more elaborate discussion of the intersection form see Section \ref{A1sec}.} 
\bea
\mathcal{C}.\can=\int\limits_X\, \omega_\mathcal{C} \wedge c_1(\can)=\int\limits_\mathcal{C}\,c_1(\can)={-1\over 2\pi} \,\int\limits_\mathcal{C}\,R_{i\bar{j}} \;i\,dz^i\wedge d\bar{z}^j
\eea
It follows that $\mathcal{C}.\can<0$ means that the Ricci tensor is on average positive on every complex curve $\mathcal{C}\subset X$, whereas $\mathcal{C}.\can>0$ means that it is negative on every complex curve $\mathcal{C}\subset X$. Hence one can say that the del Pezzo surfaces are of `positive curvature' and the surfaces of general type are of `negative curvature'.

The fact that will be important for what follows is that, if $\mathcal{C}.\can>0$, the formula (\ref{genus}) above implies for a genus-zero curve $\mathcal{C}.\mathcal{C}<-2$, whereas for $\mathcal{C}.\can<0$ it implies $\mathcal{C}.\mathcal{C}>-2$. In other words, in a positively curved surface the self-intersection number of a holomorphic sphere $\CP^1$ is larger than $-2$, whereas in a negatively curved surface it is smaller than $-2$.

(\ref{genus}) is a neat algebraic formula, but it does not provide an intuitive understanding of what is actually going on. What does provide such an intuition is the differential-geometric approach to the same issue. In fact, intersection theory tells us that the self-intersection number $\mathcal{C}.\mathcal{C}$ is the degree of the normal bundle $\mathrm{N}_{\mathcal{C}/X}$ to $\mathcal{C}$ in $X$, or in other words the integral of the first Chern class over $\mathcal{C}$:
\bea
\mathcal{C}.\mathcal{C}=\int\limits_\mathcal{C} c_1(\mathrm{N}_{\mathcal{C}/X})
\eea
Any line bundle over $\CP^1$ is a power of the Hopf bundle, i.e. $\mathcal{O}(m)$ for some integer $m$, therefore if $\mathrm{N}_{\mathcal{C}/X}=\mathcal{O}(-m)$, then $\mathcal{C}.\mathcal{C}=-m$. In particular, the $(-2)$-curves are embedded with the normal bundle $\mathcal{O}(-2)$. On the other hand, $\mathcal{O}(-2)$ is the canonical bundle of $\CP^1$, so the neighborhood of $\mathcal{C}$ in $X$ looks like the total space of the canonical bundle $\can_\mathcal{C}$. Clearly, it is a non-compact Calabi-Yau manifold (see also \textsection \,\ref{EH1}). Moreover, an explicit metric on this non-compact Calabi-Yau is known --- it is the Eguchi-Hanson metric, which we will discuss below in Sections \ref{EH1}, \ref{EH2}.\vspace{0.3cm}

\emph{Comment.} The differential-geometric approach to the intersection indices that we have just described allows to interpret the formula (\ref{genus}) as expressing the genus of a curve $\mathcal{C}\subset X$ in terms of the degree of the normal bundle to $\mathcal{C}$ and the average Ricci curvature of $\mathcal{C}$.\vspace{0.3cm} ${\scriptstyle \blacksquare}$

The fact that the Ricci form representing the canonical class of this `glued-in' manifold is zero explains the fact that blowing-up of the du Val singularities does not affect the canonical class of the ambient variety. There is yet another, perhaps simpler, way to understand this. All du Val singularities correspond to orbifolds $\CC^2/\Gamma$, where $\Gamma \subset SU(2)$ is a finite subgroup. The important fact here  is that $\Gamma$ is a finite subgroup of $SU(2)$, rather than $U(2)$. The point is that, as is well-known, the fact that the manifold is Calabi-Yau means that there exists a \underline{nowhere vanishing} \emph{holomorphic} (hence nonsingular) volume form. Of course, on $\CC^2$ with coordinates $(z_1, z_2)$ such a form is just $\Omega=dz_1 \wedge dz_2$. For this form to descend to the quotient $\CC^2/\Gamma$ any transformation $g\in \Gamma$ should have determinant $1$: $\det{g}=1$, since under the action of $g: \Omega \to \det{g} \,\cdot\,\Omega$.

\newcommand\scalemath[2]{\scalebox{#1}{\mbox{\ensuremath{\displaystyle #2}}}}
\arraycolsep=2.5pt
$A_n$ singularities of the type (\ref{A-n}) correspond to the case when $\Gamma=\mathds{Z}_{n+1}$ is a cyclic group of order $n+1$, generated by $g=\left(\scalemath{0.75}{\begin{array}{cc}e^{2\pi i \over n+1}& 0\\ 0& e^{-{2\pi i \over n+1}}\end{array}}\right)$. Indeed, the basic invariants of this action are $x=z_1^{n+1}, y=z_2^{n+1}, z=z_1z_2$, and they are related by the equation (\ref{A-n}).

\section{The surfaces}\label{surfaces}

The surfaces that will serve as the testing ground for the conjecture (\ref{Kahlerexpansion}) are the del Pezzo surface of rank $1$, termed $\mathbf{dP}_1$ (for the $\mathcal{O}(-1)$ case), and the resolved $A_1$-singularity (for the $\mathcal{O}(-2)$ case). In both cases, in order to construct the metric on these surfaces, we will use the symplectic quotient construction, the difference being that in the $\mathbf{dP}_1$ case we will use the K\"ahler quotient, and in the $A_1$-case we will use the hyper-K\"ahler quotient (as the resolved manifold turns out to be hyper-K\"ahler). Since the hyper-K\"ahler quotient is in a sense a specialization of the K\"ahler quotient, we will start with the $\mathbf{dP}_1$ case in Section \ref{dP1sec} and then turn to the $A_1$ case in Section \ref{A1sec}.

\subsection{The del Pezzo surface of rank $1$: $\mathbf{dP}_1$}\label{dP1sec}

Del Pezzo surfaces of various ranks are remarkable objects in their own right, so for general information about them we refer the reader to \cite{Manin} and \cite{Dolgachev}.

The del Pezzo surface of rank 1 has degree $d=9-1=8$. This means it can be embedded into $\CP^8$, and the embedding is given by those sections of $\mathcal{O}_{\CP^2}(3)$ which vanish at a given point on $\CP^2$, for example at $(z_1:z_2:z_3)=(0:0:1)$. We can choose the following basis for these sections of $\mathcal{O}_{\CP^2}(3)$:
\bear\label{dP1}
&& x_1=z_1^3,\;x_2=z_1^2 z_2,\;x_3=z_1^2 z_3,\;x_4=z_1 z_2^2,\;x_5=z_1 z_3^2,\;\\ \nonumber
&& x_6=z_1 z_2 z_3,\;x_7=z_2^2 z_3,\;x_8=z_2 z_3^2,\;x_9=z_2^3
\eear
These are all possible cubic monomials in 3 variables with $z_3^3$ omitted, since it is the only one that does not vanish at the prescribed point. If we now regard $(x_1, \ldots, x_9)$ as homogeneous coordinates on the projective space $\CP^8$, then the above formulas (\ref{dP1}) provide the embedding. The variables $x_1 \ldots x_9$ are not independent however, as they satisfy a wealth of algebraic equations. How do we find these equations? First of all, let us form a $3\times 9$ matrix of exponents $m=(m_{1i}, m_{2i}, m_{3i}): x_i= z_1^{m_{1i}}\,z_2^{m_{2i}}\,z_3^{m_{3i}}$. It is clear that in order to have an equation
\bea
\prod\limits_{k=1}^9 x_k^{\beta_k}=1 \quad\Rightarrow\quad \sum\limits_{i=1}^9 \beta_i \,m_{\alpha i}=0
\eea
In other words, if we regard $m$ as a map $m: \CC^9 \to \CC^3$, then $\beta \in \mathrm{Ker}(m)$. In the present case the matrix $m$ and its kernel have the following form:
\bea
m= \left(\scalemath{1}{
\begin{array}{ccccccccc}
3&2&2&1&1&1&0&0&0\\
0&1&0&2&0&1&2&1&3\\
0&0&1&0&2&1&1&2&0
\end{array}}\right),\quad\quad
\mathrm{Ker}\,m=
\left(\scalemath{1}{
\begin{array}{ccccccccc}
2&-3&0&0&0&0&0&0&1\\
2&-1&-2&0&0&0&0&1&0\\
2&-2&-1&0&0&0&1&0&0\\
1&-1&-1&0&0&1&0&0&0\\
1&0&-2&0&1&0&0&0&0\\
1&-2&0&1&0&0&0&0&0
\end{array}}\right)^{\mathrm{T}}
\eea

As a result, we observe that the variables $x_i$ satisfy the following 6 equations:
\bear\label{dP1eqs}
\begin{array}{ccc}
x_1^2 x_9=x_2^3, &\quad x_1^2 x_8=x_2 x_3^2, &\quad x_1^2 x_7=x_2^2 x_3 \\
x_1 x_6=x_2 x_3, &\quad x_1 x_5=x_3^2, &\quad x_1 x_4=x_2^2
\end{array}
\eear
It turns out that there is a different way to arrive at the same set of equations, namely $\mathbf{dP}_1$ may be constructed as a geometric invariant theory (GIT) quotient
\bea
\mathbf{dP}_1=\CC^4/(\CC^{\times})^2
\eea
For simplicity let us first elaborate on a similar construction for the affine cone over $\mathbf{dP}_1$. The \emph{affine cone} simply means that in the defining equations (\ref{dP1eqs}) we should forget about the projective equivalence relation on the variables $x_i$, i.e. we should think of them as living in $\CC^9$ rather than in $\CP^8$ \footnote{Note that, in general, the affine cone $\mathrm{Cone}(X)$ depends on the embedding of $X$ in some projective space.}. The affine cone over $\mathbf{dP}_1$ may be represented as a GIT quotient as follows:
\bea
\mathrm{Cone}(\mathbf{dP}_1)=\CC^4/\CC^{\times}
\eea
The action of $\CC^\times$ on $\CC^4$ can be diagonalized, and in this case it is given by a `charge vector' $q=(q_1, \ldots, q_4)$, so that $\CC^\times$ acts on $\CC^4$ as follows:
\bea
\CC^\times \circlearrowleft \CC^4: \quad\quad (z_1, z_2, z_3, z_4)\to (\lambda^{q_1} z_1, \lambda^{q_2} z_2, \lambda^{q_3} z_3, \lambda^{q_4} z_4)
\eea
There exists a general procedure that allows to find the charge vector(s) given a set of monomial equations, like the one in (\ref{dP1eqs}), however not to overload the exposition we will simply provide the relevant charge vector and then check that it gives rise to the correct set of equations. We claim that the charge vector in the case of $\mathbf{dP}_1$ is
\bea
q=(-2,-2,1,3)
\eea
Hence the above procedure is telling us that the affine cone over $\mathbf{dP}_1$ can be obtained as a GIT quotient
\bea
\mathrm{Cone}(\mathbf{dP}_1)=\CC^4/\CC^{\times},
\eea
where the action of $\CC^{\times}$ is given by
\bea
(z_1, z_2, z_3, z_4)\to (\lambda^{-2} z_1, \lambda^{-2} z_2, \lambda\, z_3, \lambda^3 z_4)
\eea
Now that we know the transformation property of the $z_i$'s it is easy to check the above statement directly. Indeed, construct the complete set of independent invariants
\bear
&&\tilde{x}_1=z_1^3 z_4^2,\quad \tilde{x}_2= z_1^2 z_2  z_4^2,\quad \tilde{x}_3=z_1^2 z_3 z_4,\quad \tilde{x}_4=z_1 z_2^2  z_4^2 \\ \nonumber
&&\tilde{x}_5=z_1 z_3^2,\quad \tilde{x}_6=z_1 z_2 z_3 z_4,\quad \tilde{x}_7=z_2^2 z_3 z_4,\quad \tilde{x}_8= z_2 z_3^2 ,\quad \tilde{x}_9=z_2^3 z_4^2
\eear
It is easy to see that the variables $\tilde{x}_i$, evaluated at $z_4=1$, precisely coincide with the $x_i$ coordinates (\ref{dP1}) of the del Pezzo surface. In particular, they satisfy the same system of equations (\ref{dP1eqs}). We can say that the $x_i$ are the $\tilde{x}_i$ in a particular $\CC^{\times}$-gauge. In order to pass from $\mathrm{Cone}(\mathbf{dP}_1)$ back to $\mathbf{dP}_1$ itself we should restore the projective equivalence relation on the coordinates $x_i$, which, from the point of view of the variables $z_a$, implies that we should take a further quotient by $\CC^\times$. The charge vector for this action of the second $\CC^\times$ is described in \textsection \,\ref{dP1metr}.

We will use the above quotient representation in Section \ref{dP1metr} to construct a metric on the surface $\dP_1$. In particular, we will see what the blow-up of a point means from a differential-geometric point of view.

\subsection{The SUSY $\sigma$-model setup of the K\"ahler quotient}\label{SUSYKahler}

Most of the information on the K\"ahler and hyper-K\"ahler quotient that we review and use here is contained in the paper \cite{HKLR}.

The simplest and most vivid example of K\"ahler quotient is $\CC^{N+1}//U(1)=\CP^N$. In the form, which is morally equivalent to the one discussed here, it was already used in \cite{adda}. Let us discuss this simplest example for a while.

The Lagrangian for the bosonic $\CP^N$ $\sigma$-model has the form:
\bea
\mathfrak{L}= \sum\limits_{I=1}^{N+1}\mathcal{D}_\alpha z^I\,\mathcal{D}_\alpha \bar{z}^I,\quad\textrm{where}\quad \mathcal{D}_\alpha z^I:=\dd_\alpha z^I-i A_\alpha z^I
\eea
In this formula the variables $z^I, \bar{z}^I$ are confined to the sphere $S^{2N+1}: \;$ $\sum\limits_{J=1}^{N+1} |z^J|^2=1$. Besides, $A_\alpha$ is an auxiliary gauge field --- it does not have a kinetic term, therefore it can be eliminated in an algebraic way, leading to the canonical Fubini-Study form of the Lagrangian. The introduction of this gauge field models the symplectic quotient $\mu^{-1}(0)/U(1)$, where $\mu(z, \bar{z}):=\sum\limits_{J=1}^{N+1} |z^J|^2-1$ is the moment map for the action of $U(1)$ on $\CC^{N+1}$.

On the other hand, the situation of interest to us is when $\mu^{-1}(0)/U(1)$ is not just a symplectic quotient, but rather a K\"ahler quotient. Therefore it would be desirable to work with a K\"ahler potential instead of the symplectic (K\"ahler) form. Such opportunity is precisely provided by the supersymmetric approach, due to the fact that, quite generally, the Lagrangian of a supersymmetric $\sigma$-model with K\"ahler potential $K(Z, \bar{Z})$ can be written in $\mathcal{N}=2$ (2D) superspace in a very compact form:
\bea\label{K}
\mathfrak{L}=\int\,d^2\theta\,d^2\bar{\theta}\;K(Z, \bar{Z}),
\eea
where $Z$ and $\bar{Z}$ are now chiral superfields, representing the complex coordinates on the target space $X$. Since we have assumed that $X=Y//U(1)$, we wish to write the Lagrangian using the K\"ahler potential $\tilde{K}(\tilde{Z}, \bar{\tilde{Z}})$ on $Y$ and introducing an auxiliary $U(1)$ gauge field. Suppose for simplicity that $U(1)$ acts linearly on the coordinates $\tilde{Z}, \bar{\tilde{Z}}$. In this case we introduce a gauge superfield $V$, in terms of which the Lagrangian has the form \cite{weinberg}
\bea\label{Ktil}
\mathfrak{L}'=\int\,d^2\theta\,d^2\bar{\theta}\;\left( \tilde{K}(e^V\bar{\tilde{Z}}, \,e^V\,\tilde{Z})+2 r V\right)
\eea
The second term is the Fayet-Iliopoulos term ($r$ is a real constant). Indeed, the integral $\int\,d^2\theta \, d^2\bar{\theta}\;V$ picks out the top component of $V$, i.e. the coefficient function in front of $\theta^1\,\bar{\theta}^1\,\theta^2\,\bar{\theta}^2$, which is the auxiliary field $D$. The passage from (\ref{Ktil}) back to (\ref{K}) amounts to extremizing $\mathfrak{L}'$ with respect to the vector field $V$ --- the value at the extremum is exactly $\mathfrak{L}$. Suppose $v=v_h+v_{ah}$ is the vector field describing the original $U(1)$ action on $Y$, and $v_h, v_{ah}$ are respectively the holomorphic and antiholomorphic components. Then, if $\cst$ is the complex structure on $Y$, $\cst\,v=i(v_h-v_{ah})$. The important point is that differentiation with respect to $V$ amounts to taking the derivative along $\cst v$, i.e.
\bea
\frac{\dd}{\dd V} \tilde{K}(e^V\bar{\tilde{Z}}, \,e^V\,\tilde{Z}) = \nabla_{\cst v} \,\tilde{K}(\bar{\tilde{Z}}, \,\tilde{Z})\big|_{(\tilde{Z}, \bar{\tilde{Z}})\to (e^V\tilde{Z}, e^V\bar{\tilde{Z}})}\eea
In Appendix \ref{Kahmomrel} we show that the K\"ahler potential is related to the moment map for the action of $U(1)$ by ${1\over 2}\nabla_{I\, v} \tilde{K}=\mu-r$, where $r$ is some constant. This constant is precisely provided by the Fayet-Iliopoulos term, so that the extremization of (\ref{Ktil}) gives:
\bea
\frac{1}{2}\frac{\dd}{\dd V} \tilde{K}(e^V\bar{\tilde{Z}}, \,e^V\,\tilde{Z})+r=\mu(e^V\bar{\tilde{Z}}, \,e^V\,\tilde{Z})=0
\eea
One needs to solve this equation for $V$ and then substitute it back into (\ref{Ktil}).

Using the supersymmetric language that we have just discussed, many K\"ahler metrics\footnote{In fact, most of these metrics are hyper-K\"ahler, but this is not important for us here.} on cotangent bundles to symmetric spaces were constructed in \cite{nitta}.

\subsection{The K\"ahler quotient in mathematical language}\label{Kahlermath}

In this paragraph we will discuss the mathematical theory of how to obtain the K\"ahler potential $\tilde{K}$ on the quotient $\mathcal{M}//G$ in terms of the K\"ahler potential $K$ on the original manifold $\mathcal{M}$. The corresponding ideas are outlined in \cite{HKLR}, so here we will state the results, and a direct derivation is provided in Appendix \ref{appKahquot}. As is shown in Appendix \ref{Kahmomrel}, the K\"ahler potential invariant under a group $G$ is related to the moment map $\mu$ as
\bea\label{momkahpot}
\mu_a={1\over 2}\nabla_{\cst V_a}K+r_a,
\eea
where $r_a$ are constants and $V_a$ is the vector field corresponding to Lie algebra element $a$. In fact, for simplicity we will assume that $G=U(1)^n$, so the index $a$ runs from $1$ to $n$. Now, suppose we are given an arbitrary point $p\in\mathcal{M}$. We can set up the flow equations along the force lines of the vector fields $\cst V_a$. To do this, first of all let us introduce explicitly a set of complex coordinates $z^i$ on $\mathcal{M}$. With respect to these coordinates we can split the vector fields $V_a$ in two pieces --- the holomorphic and antiholomorphic parts:
\bea
V_a=v_a+\bar{v}_a
\eea
The complex structure $\cst$ acts on this vector field as follows:
\bea
\cst V_a=i\,(v_a-\bar{v}_a)
\eea
We assume that the point $p$ itself has complex coordinates $w^1,\cdots,w^N$. So we write down the flow equations in holomorphic terms:
\bea\label{floweqs}
\frac{\dd z^i}{\dd t^a}=i\,v_a^i,\;\;\;\; \frac{\dd \bar{z}^i}{\dd t^a}=-i\,\bar{v}_a^i,\quad\quad a=1\ldots n
\eea
The initial data is that the trajectory starts at $p$, i.e. $z^i(0)=w^i,\;\;\bar{z}^i(0)=\bar{w}^i$. Next we note that at some point in time $t_a=\tau_a$ every such trajectory $z(t)$ will intersect the surface $\mu^{-1}(0)$. In other words, we define $\tau_a$ by the equation
\bea
\mu_a(z^i(\tau),\;\bar{z}^i(\tau))=0
\eea 
Hence effectively we have defined a projection
\bea
\pi: \big(w^1, \bar{w}^1, \cdots, w^N, \bar{w}^N \big) \to \big( z^1(\tau), \bar{z}^1(\tau), \cdots, z^N(\tau), \bar{z}^N(\tau) \big)
\eea
 or, more conceptually,
\bea
\pi: \mathcal{M} \to \mu^{-1}(0)
\eea
Now that we have all the necessary definitions, we can write out the K\"ahler potential $K$ on the quotient $\mathcal{M}//G$. More precisely, we will regard $K$ as a $G$-invariant function on $\mu^{-1}(0)$, which therefore descends to the quotient. The advantage of viewing it as a function on $\mu^{-1}(0)$ is that we can take a further pull-back to $\mathcal{M}$ under $\pi$, i.e. $\pi^{\ast}(K)$, to obtain:
\bea\label{Kquot}
\pi^\ast(K)=\tilde{K}(z(w, \tau),\bar{z}(\bar{w},\tau))+2\,\sum\limits_{a=1}^n\,r_a\,\tau_a(w, \bar{w})
\eea

\subsection{The K\"ahler quotient for the rank $1$ del Pezzo surface}\label{dP1metr}

We have seen above that the cone over a del Pezzo surface $\dP_1$ can be obtained as a K\"ahler quotient
$\mathrm{Cone}(\dP_1)=\mathrm \CC^4/\CC^{\times}$, where the action of $\CC^{\times}$ is given by
\bea
(\CC^{\times})_1: \;\;(z_1, z_2, z_3, z_4) \to (\lambda^{-2} z_1, \lambda^{-2} z_2, \lambda\, z_3, \lambda^3 z_4)
\eea
We are interested in the surface itself, rather than a cone over it, so we would like to take a further quotient by yet another $\CC^{\times}$ that should play a role of `projectivization' of the equations (\ref{dP1eqs}). We claim that the action of the extra $(\CC^{\times})_2$ should take the form
\bea
(\CC^{\times})_2: \;\; (z_1, z_2, z_3, z_4) \to (\eta z_1, \eta z_2, \eta z_3, z_4)
\eea
The reason for such choice is that in the $(\CC^{\times})_1$-`gauge' $z_4=1$ the action of $(\CC^{\times})_2$ is precisely the projective action on $z_1, z_2, z_3$. The moment maps for these two actions are
\bear
&&\mu_1=|z_3|^2-2(|z_1|^2+|z_2|^2)+3 |z_4|^2\\
&&\mu_2=|z_1|^2+|z_2|^2+|z_3|^2
\eear
The procedure explained in the previous Section tells us that we ought to solve the following system of equations for $(\lambda, \eta)$:
\bear\label{quoteqs1}
&&\eta \;\big(\lambda\,|z_3|^2-2\frac{|z_1|^2+|z_2|^2}{\lambda^2}\big)+3 \lambda^3\,|z_4|^2=r_1\\ \nonumber
&& \eta \;\big(\lambda |z_3|^2+\frac{|z_1|^2+|z_2|^2}{\lambda^2} \big)=r_2
\eear
To make the connection with the supersymmetric description of paragraph \ref{SUSYKahler} one should think of $\lambda$ as $e^{V_1}$ and $\eta$ as $e^{V_2}$.

Once we find $(\lambda, \eta)$, the K\"ahler potential for the quotient metric can be deduced: 
\bea
K=\eta \;\big(\lambda |z_3|^2+\frac{|z_1|^2+|z_2|^2}{\lambda^2} \big)+\lambda^3\,|z_4|^2-r_1 \log\,\lambda-r_2 \log\,\eta
\eea
If $z_3 \neq 0$ and $z_4 \neq 0$, we may choose a complete gauge $z_3=z_4=1$. Then it is clear from the above expressions that $K$ only depends on $|z_1|^2+|z_2|^2:=x$, which is a manifestation of the $U(2)$-symmetry of $\dP_1$. Using (\ref{quoteqs1}), one can simplify the expression for the K\"ahler potential, which (upon dropping an inessential constant) looks as follows:
\bea\label{kahdp1}
K=\lambda^3-r_1 \log\,\lambda-r_2 \log\,\eta
\eea
Using this explicit expression for the metric, we can deduce complete information about the Betti numbers of the surface
(see Section \ref{topology} below).

\vspace{0.3cm}
\emph{Comment.} We will only consider the case $r_1<r_2$, since this is the situation when there is a blown-up sphere at the origin. At $r_1=r_2$ the surface undergoes a topology change (the blow-down), so that for $r_1>r_2$ the sphere at the origin disappears. $\scriptstyle \blacksquare$
\vspace{-0.2cm}
\begin{center}
\line(1,0){450}
\end{center}
\textbf{Observation 1.} Expanding the K\"ahler potential (\ref{kahdp1}) in the vicinity of the blown-up $\CP^1$ (i.e. for $x\to 0$) with normal bundle $\mathcal{O}(-1)$ we obtain:
\bea
K=a_1\,\log(x)+b_1\,x+\ldots,\quad\quad a_1, b_1>0
\eea
Here $a_1$ is the K\"ahler modulus (radius squared) of the glued-in $\CP^1$, and $b_1$ is an inessential constant that can be removed by a rescaling of $x$.
\begin{center}
\line(1,0){450}
\end{center}
Despite the fact that the K\"ahler potential (\ref{kahdp1}) itself is not particularly distinguished --- there may be many other K\"ahler metrics on the same surface, --- it is a plausible assumption that the behavior at $x\to 0$ is \emph{universal}. The results presented further in the paper support this conclusion.

\subsection{Resolved ADE singularities}\label{A1sec}

In this Section we will discuss the algebraic side of the resolution of the $A_n$-singularity (\ref{A-n}). Strictly speaking, for our discussion we only need to consider $n=1$, in which case the resolved variety contains a $\CP^1$ with self-intersection number $-2$. However, in order to elucidate the algebraic side of the idea of self-intersection we also consider the case $n=2$ (higher-$n$ cases are not conceptually different). The resolution of ADE surface singularities is covered in the literature; we can recommend \cite{burban}, for instance.

\subsubsection{The $\mathds{A}_1$ singularity.}\label{a1sing}

The $\mathds{A}_1$ singularity is defined by an equation of the following form\footnote{It can be cast in the form (\ref{A-n}), $n=1$, by an obvious linear change of variables.}:
\bea\label{eqa1}
X_1:=\{\,x^2+y^2+z^2=0,\quad\textrm{where}\;\;\;(x, y, z) \in \CC^3\,\}
\eea
The l.h.s. of the equation vanishes together with its first derivatives at the origin, therefore the origin is a singular point. The blow-up corresponds to the following set of equations in $\CC^3 \times \CP^2$:
\bea\label{blowup}
xv=yu, \quad xw=zu, \quad yw=zv,
\eea
where $(u:v:w)$ is a set of homogeneous coordinates on a new $\CP^2$. Intuitively these equations mean that the vector $(u, v, w)$ is proportional to the vector $(x, y, z)$, if the latter is nonzero. However, the variables $(u, v, w)$ live in projective space, hence they only capture the angular direction of the vector $(x, y, z)$ and not its modulus. Therefore the singularity is `resolved' in the following fashion: by means of (\ref{blowup}) we keep track of its fine structure --- now the angle at which we approach the singular point is taken into account. More exactly, we lift the surface $X_1$ to $\CC^3\times \CP^2$:
\bea\label{tildeX}
\tilde{X}_1:=\textrm{The closure of }\, \left\{\,\begin{array}{c}x^2+y^2+z^2=0 \\ \;xv=yu, \; xw=zu, \; yw=zv \end{array}\quad\textrm{in}\;\;\;\CC^3/\{0\}\; \times\; \CP^2\,\right\}
\eea
Let us say a few words about this definition. First of all, the derivatives of the equations in (\ref{tildeX}) no longer vanish at the origin, so the singularity is removed. Defining a projection $\pi: \tilde{X}_1 \to X_1$ that `forgets' the $(u:v:w)$ coordinates, we find that in this language the singularity of $X_1$ may be seen as the singularity of the projection map. Secondly, formally speaking, one solution to the equations in (\ref{tildeX}) is $x=y=z=0, \; (u:v:w)$ arbitrary. This is a copy of $\CP^2$. However, in the definition of $\tilde{X}_1$ we do not want to include the whole of this $\CP^2$. We rather wish to include only those points of this $\CP^2$ that form the closure of the rest of the surface as $(x, y, z)\to 0$. This is the reason that we first exclude the point $x=y=z=0$ and then add all limiting points $\overline{\pi^{-1}(X_1/\{0\})} \cap \CP^2$ (the closure). We will now see that these limiting points form a $\CP^1$, which is usually referred to as the exceptional divisor of the blow-up. This is the holomorphic sphere that we are after.

One needs to consider three charts, in which $u=1, v=1$ or $w=1$ respectively. The three cases are analogous and for simplicity we will consider the case $u=1$. Then $y=xv$ and $z=xw$, so
\bea
x^2+y^2+z^2=0 \quad\Rightarrow \quad x^2(1+v^2+w^2)=0
\eea
If $x=0$, then $y=z=0$, which leaves the $\CP^2$ glued in at the origin. On the other hand, the equation $1+v^2+w^2=0$ in the other two charts looks as $u^2+1+w^2=0$ and $u^2+v^2+1=0$, so we can restore the homogeneous coordinates and write it as $\Sigma:=\{u^2+v^2+w^2=0\}$. Hence it is a quadric in $\CP^2$, and, as such, it is isomorphic to $\CP^1$ --- the isomorphism is given by the Veronese map, described after formula (\ref{A-n}). We wish to show that $\Sigma$ has self-intersection index $-2$.

Calculating the self-intersection number $\mathcal{C}.\mathcal{C}$ of a curve $\mathcal{C} \subset X$ is a rather tricky enterprise. Indeed, when two submanifolds are distinct and of complementary dimensions, it is clear what their intersection number means. It is less obvious what it means to calculate the intersection number of a submanifold with itself. The idea here is that we should think of the self-intersection in a somewhat homological way. Indeed, suppose $\mathcal{C}_{1}, \mathcal{C}_2\subset X$ are two smooth complex curves. Via Poincare duality, we can associate to them the 2-forms $\omega_1, \omega_2 \in H^2(X, \mathds{Z})$, for which the intersection number is the simple integral $\mathcal{C}_1.\mathcal{C}_2:=\int\limits_X \omega_1 \wedge \omega_2$. The r.h.s. is well-defined for $\omega_1=\omega_2:=\omega$ as well, i.e. we can define the self-intersection $\mathcal{C}.\mathcal{C}:=\int\limits_X \omega \wedge \omega$. What this really means is that the self-intersection is the intersection number of two submanifolds $\mathcal{C}_1$ and $\mathcal{C}_2$ that are in the same homology class, but have a transverse intersection\footnote{`Transverse' means that at each intersection point $p$: $\mathrm{T}_p\mathcal{C}_1 \oplus \mathrm{T}_p \mathcal{C}_2= \mathrm{T}_p X$. Clearly, in this case one can assign a well-defined intersection index at the point $p$.}.  Analogously to the topological situation, calculating the self-intersection number $\mathcal{C}.\mathcal{C}$ algebraically means calculating the intersection number of $\mathcal{C}$ with some $\tilde{\mathcal{C}}$ that is `\emph{in the same class}' with $\mathcal{C}$. In the context of homology, `in the same class' means `in the same homology class'. In the algebraic setting, however, `in the same class' means that the divisor should be in the same `linear equivalence class', that is $\tilde{\mathcal{C}}=\mathcal{C}+D_f$, where $D_f$ is the divisor of some function $f$. This follows from the important theorem that the intersection indices are invariant under the  addition of divisors of functions \cite{Shafarevich}, i.e. for any divisor $\mathcal{E}$: $\mathcal{E}.D_f=0$. The function $f$ should be chosen in such a way that it has a pole at $\mathcal{C}$, then its divisor includes a term $-\mathcal{C}$, so that $\tilde{\mathcal{C}}$ no longer contains $\mathcal{C}$ as its component, and therefore it will be possible to compute the intersection $\tilde{\mathcal{C}}.\mathcal{C}$, that is equal to the self-intersection number $\mathcal{C}.\mathcal{C}$. 

To calculate the self-intersection number $\Sigma.\Sigma$, we choose a function $f=\frac{1}{x}$ and calculate its divisor. $x=0$ implies either $y=z=0$, in which case we arrive back at $\Sigma$, or $u=0$, then we have the remaining equations
\bea
\{ \;y^2+z^2=0,\quad\quad yw=zv\;\} \in \CC^2 \times \CP^1
\eea
The two solutions are $y=\pm i z, \;(0:\pm i:1)\in \CP^2$. We call these two affine lines $L_1, L_2$. Hence the divisor of $f$ is
\bea
D_f=-(\Sigma+L_1+L_2)
\eea
and the self-intersection $\Sigma.\Sigma$ is
\bea
\Sigma.\Sigma:=\Sigma.\big(\Sigma+D_f \big)=-\Sigma.L_1-\Sigma.L_2
\eea
Since $\Sigma$ is located at $x=y=z=0$, it intersects $L_1$ and $L_2$ at the points $y=z=0$, $(0:\pm i:1)$, so $\Sigma.L_1=\Sigma.L_2=1$, therefore
\bea
\Sigma.\Sigma=-2
\eea

\subsubsection{The $\mathds{A}_2$ singularity.}\label{a2sing}

The $\mathds{A}_2$ singularity is defined by an equation of the following form:
\bea\label{A2eq}
x^2+y^2+z^3=0
\eea
The blow-up is described by the same equations (\ref{blowup}) as in the previous example. As before, we consider three charts covering the $\CP^2$ with homogeneous coordinates $(u:v:w)$.

\vspace{0.3cm}
1. $u=1 \Rightarrow y= x v, z= x w.$ Substituting into (\ref{A2eq}), we obtain $x^2 (1+ v^2+x w^3)=0$, which implies either $x=y=z=0, \;(u:v:w)$ arbitrary, or $1+v^2+x w^3=0$. In order to find the closure of this latter surface we simply set $x=0$, obtaining $v=\pm i$. Therefore the image of the exceptional divisor in this coordinate chart consists of two copies of $\CC$, namely $(1 : \pm i : w)$. We will now see that, in a different chart, corresponding to $w\to\infty$, these two copies of $\CC$ acquire an additional point, common to both of them, therefore the exceptional divisor consists, in fact, of two copies of $\CP^1$ intersecting (transversely) in one point.\vspace{0.5cm}

2. $v=1 \Rightarrow x= y u, z= y w.$ Substituting into (\ref{A2eq}), we obtain $y^2 (1+ u^2+y w^3)=0$, which implies either $x=y=z=0, \;(u:v:w)$ arbitrary, or $1+ u^2+y w^3=0$. The closure of this latter surface consists of two copies of $\CC$, once again, with coordinates $(\pm i:1:w)$. Since the coordinates on $\CP^2$ may be rescaled by $\pm i$, these are exactly the same copies of $\CC$ as the ones in the $u=1$ chart. So in this chart we have obtained nothing new.\vspace{0.5cm}

3. $w=1 \Rightarrow y= z v, x= z u.$ Substituting into (\ref{A2eq}), we obtain $z^2 (v^2+ u^2+z)=0$, which implies either $x=y=z=0, \;(u:v:w)$ arbitrary, or $v^2+ u^2+z=0$. The closure of the latter surface is given by $v=\pm iu$, i.e. in this patch we obtain two copies of $\CC$ given by $(u:\pm i u: 1)$. These are the same of copies of $\CC$ that we encountered above, but viewed in a different coordinate patch. Moreover, in this patch they intersect, at $u=0$, and the intersection is transverse.\vspace{0.5cm}

Hence, we have found two copies of $\CP^1$, that we will refer to as $\Sigma_\pm$ with the intersection $\Sigma_+.\Sigma_-=1$. Just like in the $\mathds{A}_1$ case above, the most delicate part is to find the self-intersection number of each of these $\CP^1$'s.
\begin{figure}
\centering
\includegraphics[width=0.4\textwidth]{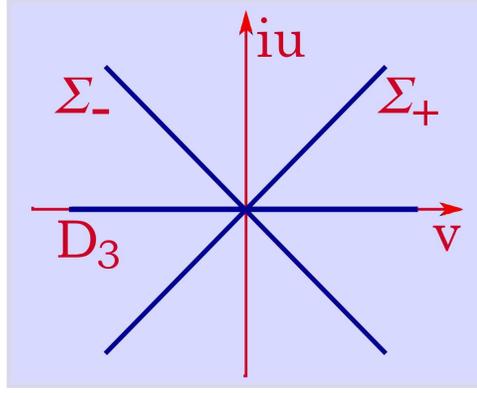}
\caption{The transverse intersection of three lines: $\Sigma_+$, $\Sigma_-$ and $D_3$.}
\label{fig:awesome_image}
\end{figure}

In the case at hand we choose an auxiliary function $f={1\over x}$. It has a pole wherever $x$ has a zero. Setting $x=0$ and examining the equations (\ref{blowup}), we find two possibilities:

(1) $x=y=z=0$, which implies that $x$ has a zero at the exceptional divisor $\Sigma_+ + \Sigma_-$

(2) $u=0$ and $\textrm{the closure of }\;\{\;y^2+z^3=0, \; yw=zv\;\} \in \CC^2/\{0\}\times \CP^1$. It is easy to see that the latter equations define a copy of $\CC$. Indeed, set $w=1$, for instance. Substituting $y=zv$ into $y^2+z^3=0$ and assuming $z\neq 0$, we obtain $z=-v^2$. Taking the closure is equivalent to setting $v=0$, i.e. including one of the poles of the $\CP^1$. The opposite pole, $w=0$, is not reached, since it corresponds to $y, z\to\infty$. Let us denote this copy of $\CC$ by $D_3$. Then
\bea
D_f=-\Sigma_+-\Sigma_- - D_3 \;\; \Rightarrow \;\; \tilde{\Sigma}_+:=\Sigma_+ + D_f=-\Sigma_- - D_3
\eea
As we discussed above, $\tilde{\Sigma}_+$ is linearly equivalent to $\Sigma_+$, therefore
\bea\label{selfint}
\Sigma_+ . \Sigma_+ =\Sigma_+ . \tilde{\Sigma}_+=-\Sigma_+ . \Sigma_--\Sigma_+. D_3
\eea
Thus the problem is reduced to the calculation of the intersection of $\Sigma_+$ with distinct divisors (curves). We have already found that $\Sigma_+ . \Sigma_-=1$. Hence what is left is to calculate $\Sigma_+. D_3$. The inspection of paragraphs (1., 2., 3.) above, which describe the exceptional divisor in the three patches, shows that the intersection with $D_3$ occurs in the 3-rd patch, when $w=1, \;u=v=0$, i.e. at the point (0 : 0 : 1). At this point $D_3$ intersects both $\Sigma_+$ and $\Sigma_-$. The intersection is transverse, since in the $(u, v)$ frame the equations of $\Sigma_\pm$ and $D_3$ are:
\bea
\Sigma_\pm:\quad v=\pm i u,\quad\quad D_3:\quad u=0
\eea
Therefore $\Sigma_+. D_3=\Sigma_-. D_3=\Sigma_+. \Sigma_-=1$. From (\ref{selfint}) it then follows that $\Sigma_+.\Sigma_+=-2$, and analogously $\Sigma_- . \Sigma_-=-2$. The intersection matrix is the Cartan matrix of $\mathds{A}_2$:
\bea
\Sigma_\pm.\Sigma_\pm=\left(
\begin{array}{cc}
 -2 & 1 \\
 1 & -2
 \end{array}
\right)
\eea

\subsection{$\mathcal{O}(-2)$ bundle over $\CP^1$. The direct construction}\label{EH1}

As we have discussed, the resolution of an $\mathds{A}_1$ singularity leads to a new variety $\tilde{X}_1$ that has a $\CP^1$ glued in at the origin with self-intersection $-2$, hence normal bundle $\mathcal{O}(-2)$. The so-called Eguchi-Hanson `gravitational instanton' is the metric on the total space of the line bundle $\mathcal{O}(-2)$ over $\CP^1$, which means it can be thought of as a metric on $\tilde{X}_1$, a noncompact complex surface.

The Eguchi-Hanson metric is the simplest example of Calabi-Yau metric one can obtain on the total space of the canonical bundle. As the base space we take $\CP^1$, therefore the canonical bundle $K=\mathcal{O}(-2)$. According to the discussion in \textsection\,\ref{a1sing}, we are looking for a Calabi-Yau metric on the affine cone
\bea
V=\{\;x^2+y^2+z^2=0\;\}\subset \CC^3
\eea
In this particular case this is not only a Calabi-Yau metric, but also a hyper-K\"ahler one. For a manifold of arbitrary dimension this would be a stronger statement, since generally the holonomy groups are related as $Sp(n) \subset SU(2n)$, however for surfaces this is the same thing, because $Sp(1)\simeq SU(2)$.

The homogeneous coordinates on $\CP^1$ are denoted $(z_1, z_2)$. They are in fact sections of the line bundle $\mathcal{O}(1)$ over $\CP^1$. Since we are interested in the total space of the canonical bundle $\mathcal{O}(-2)$, we introduce a coordinate $u$ in the fiber. As usual, $\CP^1$ can be covered by two patches with a complex coordinate $z$ in one of them, $z'$ in the opposite one and a holomorphic transition function $z'={1\over z}$ in the vicinity of the equator. If $u$ is the coordinate in the fiber over the first patch and $u'$ the coordinate in the fiber over the second patch, they are related as follows:
\bea\label{transform}
z'=\frac{1}{z},\quad u'=\frac{1}{(z')^2}\,u
\eea
In particular, the 2-form
\bea\label{holform}
\omega=dz \wedge du=-dz'\wedge du'
\eea
is holomorphic and nonzero in both patches, which implies that the canonical bundle is trivial.

\vspace{0.3cm}
\emph{Comment.} Below we will also consider the total space of the line bundle $\mathcal{O}(-m)$ over $\CP^1$ for $m\geq 3$. In this case the transition functions are
\bea\label{transform2}
z'=\frac{1}{z},\quad u'=\frac{1}{(z')^m}\,u,
\eea
hence the form $\omega_m=dz \wedge du$ in the opposite patch looks as $\omega_m=-(z')^{m-2} \,dz'\wedge du'$. Therefore it has a zero of order $m-2$ at $F=\{z'=0, \;\;u'\;\textrm{arbitrary}\}$. Hence the canonical divisor is $\mathcal{K}=(m-2)\;(F)$, and the first Chern class is $c_1(\mathcal{K})=(m-2) \,\omega_{\mathds{FS}}$, where $\omega_{\mathds{FS}}$ is the Fubini-Study form on the underlying $\CP^1$. ${\scriptstyle \blacksquare}$

\vspace{0.3cm}
From (\ref{transform}) it follows that the fiber coordinate $u$ transforms opposite to the quadratic combinations of the homogeneous coordinates $(z_1, z_2)$, i.e. $z_1^2, z_2^2, z_1z_2$. Therefore the real combinations of $z$'s and $u$ which are unchanged are ${|z_1|^4\, |u|^2}, {|z_1z_2|^2 \,|u|^2}, {|z_2|^4\, |u|^2}$. We will restrict ourselves to the metrics with $SU(2)$ isometry, where the $SU(2)$ rotates $(z_1, z_2)$ in the standard way, as a doublet. In this case the only possible combination is ${(|z_1|^2+|z_2|^2)^2\, |u|^2}$. Hence we look for a K\"ahler potential of the form
\bea\label{EHansatz}
K=K\left({(|z_1|^2+|z_2|^2)^2\, |u|^2}\right).
\eea
In this formula $K$ is a function yet to be determined from the condition of Ricci-flatness of the resulting K\"ahler metric. In imposing this condition, the important fact which drastically simplifies the calculations is that for an arbitrary K\"ahler metric $g_{i\widebar{j}}$ its Ricci tensor $R_{i\widebar{j}}$ is Hermitian, and is explicitly given by
\bea
R_{i\widebar{j}}=-\dd_i\widebar{\dd}_j\log\det\|g_{m\widebar{n}}\|
\eea
It follows that, in particular, it defines a closed 2-form. If this ought to be zero, $\det\|g_{m\widebar{n}}\|$ has to be a product of a holomorphic and an anti-holomorphic function.

We pass to the inhomogeneous coordinates by setting $u=1$. It then follows from (\ref{EHansatz}) that $K$ is a function of $x\equiv |z_1|^2+|z_2|^2$. The metric then looks as
\bea\label{EHmetric}
g=\left(\begin{array}{cc}
K'+|z_1|^2 K'' & \bar{z}_2 z_1 K''  \\
z_2 \bar{z}_1 K'' & K'+|z_2|^2 K''  \end{array}\right )
\eea
and the Ricci flatness condition $\det\|g_{m\widebar{n}}\|=|\sigma(z)|^2$ gives
\bea
K' (K' x)'=|\sigma(z)|^2,
\eea
which can only hold for $|\sigma(z)|^2=\textrm{const.}=\tilde{A}$. In this case
\bea\label{kahEH1}
K'=c\,\frac{\sqrt{r^2+4 x^2} }{x}\quad\Rightarrow\quad K=\sqrt{r^2+4 x^2}+r\, \log \left(\frac{\sqrt{
   r^2+4 x^2}-r}{2x}\right),\;\;\;r>0
\eea
where in the last formula we have set the overall scale of the metric equal to one: $c=1$. This is all one needs to recover the Eguchi-Hanson metric using the formula (\ref{EHmetric}).

\subsection{$\mathcal{O}(-2)$ bundle over $\CP^1$. The hyper-K\"ahler quotient.}\label{EH2}

As it follows from \textsection \,\ref{SUSYKahler} and \textsection\,\ref{Kahlermath} above, a single K\"ahler quotient $\tilde{X}:=Y//U(1)$ is a complex manifold with $\mathrm{dim}_\mathbb{C} \tilde{X}=\dim_\mathbb{C} Y-1$. The hyper-K\"ahler quotient that we will denote as $X:=Y\underset{\mathrm{H}}{//}U(1)$ produces from a hyper-K\"ahler manifold $Y$ a new hyper-K\"ahler manifold $X$ with $\mathrm{dim}_\mathbb{H} X=\dim_\mathbb{H} Y-1$, or in other words $\mathrm{dim}_\mathbb{C} X=\dim_\mathbb{C} Y-2$.

The hyper-K\"ahler quotient is constructed using a triplet of moment maps $\vec{\mu}$, which are conveniently organized into a real one $\mu_\mathbb{R}:=\mu_3$ and a complex one $\mu_\mathbb{C}:=\mu_1+i \mu_2$. Furthermore, $\mu_\mathbb{C}$ is not only a complex-valued, but also a holomorphic function on $Y$.

The simplest asymptotically locally Euclidean (ALE) manifold may be obtained as a hyper-K\"ahler quotient of the flat space $\mathbb{R}^8=\mathbb{C}^4=\mathbb{H}^2$, whose K\"ahler potential $K_0$ is a simple quadratic function
\bea
K_0=|a|^2+|b|^2+|c|^2+|d|^2
\eea
The moment maps defining the hyper-K\"ahler quotient are:
\bear
&&\mu_\mathbb{C}= ad-bc+\zeta\\
&&\mu_\mathbb{R}=|a|^2+|b|^2-|c|^2-|d|^2+r
\eear
The hyper-K\"ahler quotient $X$ may be seen as a usual K\"ahler quotient of the space $\mu_\mathbb{C}^{-1}(0)$, i.e.
\bea
X=\mu_\mathbb{C}^{-1}(0)//U(1)
\eea
This means that, first of all, we need to restrict $K_0$ to $\mu_\mathbb{C}^{-1}(0)$, and then carry out the K\"ahler quotient construction for the $U(1)$ action $(a, b, c, d)\to (e^{i\theta} \,a, e^{i\theta} \,b, e^{-i\theta} \,c, e^{-i\theta} \,d)$. As a result we obtain the K\"ahler potential on $X$
\bea\label{kahEH}
K_X=\lambda \big(|a|^2+|b|^2 \big)+{1\over \lambda}\big(|c|^2+|d|^2 \big)+r \log{\lambda},
\eea
where $\lambda$ is found from the equation
\bea\label{eqEH}
\mu_\mathbb{R}(\lambda a, \lambda b, \lambda^{-1} c, \lambda^{-1} d)=\lambda \big(|a|^2+|b|^2 \big)-{1\over \lambda} \big(|c|^2+|d|^2 \big)+r=0
\eea
The Eguchi-Hanson space is obtained for $\zeta=0$. In this case $\mu_\mathbb{C}= ad-bc=0$ can be solved as $d=\xi \,b, c=\xi \,a$, where $\xi$ is a new complex coordinate. One then finds that the K\"ahler potential $K_X$ is a function of $|\xi|^2 \big( |a|^2+|b|^2\big)^2$ (up to an addition of an inessential ${r\over 2}\log(\xi \bar{\xi})$) and, as such, is degenerate under the transformation $\xi \to \kappa^2 \xi, a\to \kappa^{-1} \xi, b\to \kappa^{-1}\xi$. We can remove this degeneracy by imposing a `gauge' $\xi=1$. Denoting $x:=|a|^2+|b|^2$ and rescaling $\lambda \to {\lambda\over x}$ we can rewrite the K\"ahler potential (\ref{kahEH}) and the equation (\ref{eqEH}) in the following form:
\bear
&&K_X=\lambda+{x^2\over \lambda}+r \log{\lambda\over x},\quad \textrm{where}\;\lambda\; \textrm{satisfies}\\
&&\lambda -{x^2\over \lambda} +r=0
\eear
Solving for $\lambda$ and substituting into the expression for the K\"ahler potential we find that it is exactly the same as (\ref{kahEH1}) obtained above by direct solution of the Ricci-flatness equation.
\begin{center}
\line(1,0){450}
\end{center}
\textbf{Observation 2.} Expanding the K\"ahler potential (\ref{kahEH1}) in the vicinity of the blown-up $\CP^1$ (i.e. for $x\to 0$) with normal bundle $\mathcal{O}(-2)$ we obtain:
\bea
K=a_2\,\log(x)+b_2\,x^2+\ldots,\quad\quad a_2, b_2>0
\eea
Here, just like in \textbf{Observation 1}, $a_2$ is a K\"ahler modulus of the glued-in sphere, and $b_2$ is an inessential constant (the only important fact is that it is positive).
\begin{center}
\line(1,0){450}
\end{center}
Comparing  \textbf{Observation 1} to  \textbf{Observation 2}, we arrive at a natural conjecture that, in the $U(2)$-invariant case, if a $\CP^1$ is embedded in a complex surface $X$ with normal bundle $\mathcal{O}(-m)$, then the local behavior of the K\"ahler potential near the $\CP^1$ ($x\to 0$) is as follows:
\bea
K=a_m\,\log(x)+b_m\,x^m+\ldots, \quad\quad a_m, b_m>0
\eea

If a $\CP^1$ is embedded in $X$ with normal bundle $\mathcal{O}(-m)$, then its neighborhood looks as the total space of the bundle $\mathcal{O}(-m)$. In Section \ref{KEmetr} we will see that it is possible to build negatively curved K\"ahler-Einstein metrics on these spaces for $m\geq 3$. These solutions will therefore be generalizations of the Eguchi-Hanson metric, which corresponds to $m=2$.

\subsection{The supersymmetric setup in 6D}\label{SUSY6D}

In Section \ref{SUSYKahler} we already mentioned that the K\"ahler quotient is naturally `built-in' certain supersymmetric $\sigma$-model constructions. In principle, in a similar way the hyper-K\"ahler quotient can be obtained from an analogous 2D $\sigma$-model with $\mathcal{N}=(4,4)$ supersymmetry. However, there is yet another realm in the supersymmetric domain where the hyper-K\"ahler quotient appears. This is on the so-called Higgs branch of 4D $\mathcal{N}=2$ theories or, in a slightly simpler way, in $\mathcal{N}=(1,0)$ theories in 6D. This other realm has a direct physics interpretation, namely the corresponding supersymmetric theory arises as an effective theory for the fluctuations of a six-simensional D-brane located at an $ADE$-singularity of a transverse $K3$-surface (Calabi-Yau space) \cite{DM}.

The notation $\mathcal{N}=(1,0)$ refers to the fact that the supercharge $\mathcal{Q}_{A}^a$ is a symplectic-Majorana spinor (we are talking about Minkowski signature here). Here the indices take the following values: $A=1, 2$ and $a=1, \ldots, 4$.  The supercharge satisfies a reality property $\bar{\mathcal{Q}}^B_b=\epsilon^{BA}\,C_{ba}\,\mathcal{Q}_A^a$, where $\epsilon$ is the two-dimensional $\epsilon$-tensor, and $C$ is the charge conjugation matrix. The supersymmetry algebra has the following form:
\bea
\{ \mathcal{Q}_A^a, \bar{\mathcal{Q}}^B_b \}=2i\;\delta_A^B\;(\sigma^\mu P_\mu)_a^b
\eea

The crucial fact is that this algebra has $SU(2)$ R-symmetry (in contrast to $U(1)$ R-symmetry of $\mathcal{N}=1$ SUSY algebra in 4D), which rotates the indices $A, B$: $\mathcal{Q}_A^a \to g_{AB} \mathcal{Q}_B^a$. It is nontrivial that this transformation is compatible with the reality property of the supercharge.

Such $\mathcal{N}=(1,0)$ SUSY theory has two types of multiplets: the vector multiplet (in the adjoint representation of the gauge group) and the hypermultiplet (in an arbitrary representation of the gauge group). The Lagrangian is completely determined by the field content, i.e. by the representations of the hyper-multiplets (in particular, there is no superpotential).

A stack of $N$ six-dimensional branes in flat space $\mathbb{R}^6 \times \CC^2$ ought to be described by $U(N)$ gauge theory with a single hypermultiplet, since the hypermultiplet contains $2$ complex scalars, which parametrize the normal directions to the brane. Imagine now that we wish to consider an orbifold $\CC^2/\Gamma$ instead of $\CC^2$ and place the branes at the fixed point of the orbifold. According to \cite{DM}, in order to take the orbifold quotient, one needs to specify not only $\Gamma$, but also its representation $R$ on the Chan-Paton vector space (i.e. $\CC^N$ for gauge group $U(N)$). Once this data is provided, the spectrum of fields describing the low-energy theory on the brane can be obtained by means of the projection $\Pi_g(\phi)=\phi$ for all $g\in \Gamma$, where the operator $\Pi$ is defined on the hypermultiplet as $\Pi_g(\Phi_A)=g_{AB} R(g)\Phi_B R(g)^{-1}$ and on the vector multiplet as $\Pi_g(A_\mu)=R(g)A_\mu R(g)^{-1}$. Here $g_{AB}$ is the defining representation of $\Gamma$ (or, in other words, the way $\Gamma$ acts on $\CC^2$), and the matrices $R(g)$ act on the gauge indices. In the case when $R$ is the regular representation ($\mathrm{dim}\, R=|\Gamma|$), i.e. the representation of $\Gamma$ on its group algebra, then one says that one `full' brane (as opposed to a `fractional' brane) has been placed at the singularity.

To review the supersymmetric setup for the construction of the Eguchi-Hanson space we will consider $\Gamma=\mathbb{Z}_2$, $R$ --- its regular representation, of dimension $2$. The group algebra is generated by $(1, h)$ with the relation $h^2=1$. In the basis $(1+h, 1-h)$, $1$ is a $2\times 2$ identity matrix $\Id_2$, whereas $h=\left( \begin{array}{cc}
1 & 0\\
0 & -1 \end{array} \right)$. The projection conditions then leave the components $A_\mu^{11}, \,A_\mu^{22}$ of the vector field and $\Phi_A^{12},\,\Phi_A^{21}$ of the hypermultiplet. Hence, the gauge group reduces to $U(1)^2$. However the diagonal $U(1)$ describes a free vector multiplet and decouples from the rest. The original hypermultiplet splits into two hypermultiplets with opposite charges with respect to the remaining $U(1)$. As discussed below, this is precisely the right data to obtain the Eguchi-Hanson space via the supersymmetric construction.

We will not attempt to write out the full SUSY Lagrangian, since this is a rather tedious exercise and to a large extent unnecessary for our present purposes. Instead, we will review how the triplet of auxiliary $D$-fields comes about in this 6D theory, since this is a crucial ingredient in the construction. Recall that in 4D the gauge field strength superfield $W$ is a spinor superfield, which is chiral in two senses: both as a chiral superfield (i.e. a short multiplet) and as a chiral spinor. It has the following form \cite{weinberg}:
\bea
W_L=\lambda_L(x_+)+{1\over 2} \gamma^\mu \gamma^\nu \theta_L f_{\mu\nu}(x_+)+\big(\theta_L^{\mathrm{T}}\epsilon \theta_L\big)\; \gamma^\alpha \dd_\alpha\lambda_R(x_+)-i \theta_L D(x_+)
\eea
In six dimensions the gauge field strength superfield $(W_a)_A$ is of opposite chirality to the supercharge and satisfies the following conditions:
\bear
&&(\epsilon \sigma_n)^{AB}\,\big(\mathcal{D}^a_A\,(W_b)_B- {1\over 4} \delta^a_b \;\mathcal{D}^c_A\,(W_c)_B\big)=0,\quad \quad n=1, 2, 3\\
&&\epsilon^{AB} \,\mathcal{D}^c_A\,(W_c)_B=0\\
&&\textrm{and the reality property}\;\; (W^B_b)^\ast=\epsilon^{BA}\,C_{ba}\,(W_a)_A
\eear 
These equations may be solved, and as a result we obtain the following expression for $W$, up to second order in $(\theta_a)_A$\footnote{Round brackets $(bc)$ denote symmetrization, and square brackets $[bc]$ denote antisymmetrization.}:
\bear
W^a_A=\lambda^a_A+F^{ab}\theta_A^b+(D_i \sigma_i)_{AB}C^{ab} \theta^b_B+Q^{a(bc)} \epsilon_{BC} \theta^b_B \theta^c_C+\epsilon_{MN} S^{a[bc]}_M \theta_N^b \theta_A^c+\ldots\, ,
\eear
where
\bear
&&Q_A^{a(nd)}={1\over 4} \big( C_{an} (\sigma_\mu)_{md} \dd_\mu \lambda_A^m+C_{ad} (\sigma_\mu)_{mn} \dd_\mu \lambda_A^m\big)\\
&&S_A^{a[nd]}={1\over 2} \big( C_{ad} (\sigma_\mu)_{mn} \dd_\mu \lambda_A^m-C_{an}(\sigma_\mu)_{md} \dd_\mu \lambda_A^m\big)-(C\sigma_\mu)_{nd} \dd_\mu \lambda_A^a
\eear
and the matrix of fields $F^{ab}$ is constrained to be `traceless': $C_{ab} F^{ab}=0$. The reality property for the symplectic-Majorana spinor $W^a_A$ translates into the reality properties of the component fields. In particular, $\lambda^a_A$ is a symplectic-Majorana spinor as well, $(F_{ab})^\ast$ is linearly related to $F_{ab}$, hence it has only $15$ real components, which can be packed into a skew-symmetric real-valued tensor $F_{\mu\nu}$, and $D_i$ are a triplet of real auxiliary fields. It is precisely the appearance of this triplet, in place of a singlet $D$, that is important for us here. Assuming that the theory includes $M$ hypermultiplets with scalar components $_m\phi_A^a$, where $a$ is an $U(N)$ gauge index and $m$ labels the hypermultiplet ($m=1\ldots M$), let us now write out the part of the supersymmetric Lagrangian, where the $D_i$ fields enter:
\bea\label{6DLagr}
\mathfrak{L}\sim {1\over 2} D_i^2+ D_i\,\big[ \sum\limits_{m=1}^M\; (_m\phi_A^a)^\ast (\sigma_i)_{AB} \;_m\phi_B^a +\zeta_i \big],
\eea
where $\zeta_i$ is a triplet of Fayet-Iliopoulos terms. The possibility of adding them without destroying supersymmetry comes from the fact that the variation of $D_i$ under a supersymmetry transformation is a full derivative:
\bea
\delta D_i=-{1\over 4} (\sigma_i)_{AB} \bar{\mu}_A \sigma_\mu C \dd_\mu \lambda_B
\eea
The fields $D_i$ are auxiliary, in the sense that they have no kinetic terms, so they can be integrated out of (\ref{6DLagr}) to produce
\bea
\mathfrak{L}\sim \frac{1}{2}\big[ \sum\limits_{m=1}^M\; (_m\phi_A^a)^\ast (\sigma_i)_{AB} \;_m\phi_B^a +\zeta_i \big]^2
\eea
The locus of points in field space where this function reaches a (zero) minimum is given by the hyper-K\"ahler moment map equations $\mu_i=0,\;\;i=1,2,3$. Since field configurations related by gauge transformations are equivalent, we need to take the quotient with respect to the gauge group $U(N)$, hence the space of physical field configurations saturating the minimum of the potential is the hyper-K\"ahler quotient $\{\;\mu_i^{-1}(0),\;i=1, 2, 3\;\} / U(N)$. The construction of the Eguchi-Hanson space in Section (\ref{EH2}) was a special case when $N=1$ and $M=2$.

\section{Topology of surfaces}\label{topology}

Suppose a compact complex surface $X$ is simply-connected, in this case $H^1(X, \mathbb{Z})=0$. Poincare duality implies $H^0(X, \mathbb{Z})=H^4(X, \mathbb{Z})=\mathbb{Z}$ and $H^1(X, \mathbb{Z})=H^3(X, \mathbb{Z})=0$, therefore the Euler characteristic is $\mathrm{Eu}(X)=2+\mathrm{dim}\, H^2(X, \mathbb{Z})$. Apart from the Euler characteristic, there is another important invariant, which characterizes the topology of $X$ --- it is the signature $\mathrm{Sgn}(X)$. It arises from the intersection form on $H^2(X, \mathbb{Z})$, i.e. if $\alpha, \beta \in H^2(X, \mathbb{Z})$, we can compute the `intersection number' $(\alpha, \beta):=\int\limits_X\;\alpha\wedge \beta$. It is a quadratic form on the vector space $H^2(X,\mathbb{R})$. $\mathrm{Sgn}(X)$ is by definition the signature of this quadratic form, i.e. if $n_\pm$ are the numbers of positive/negative eigenvalues of $(\,,\,)$, then
\bea
\mathrm{Sgn}(X)=n_+-n_-
\eea
The Euler characteristic, in turn, is\footnote{We assume there are no zero eigenvalues.}
\bea
\mathrm{Eu}(X)=2+n_++n_-
\eea
In the noncompact case the formula for the Euler characteristic is modified. Indeed, the noncompact surfaces that we will encounter have $H^3(X, \mathbb{Z})=H^4(X, \mathbb{Z})=0$, hence $\mathrm{Eu}(X_\textrm{noncompact})=1+n_++n_-$.

\vspace{0.3cm}
\emph{Comment.} On a complex surface $X$ there is a decomposition of the complexified second cohomology group: $H^2(X,\CC)=H^{(2,0)}(X)\oplus H^{(1,1)}(X) \oplus H^{(0,2)}(X)$, where $H^{(i,j)}$ are the Dolbeault cohomology groups. The intersection form then splits into an intersection form on $H^{(1,1)}(X)$  and a pairing between $H^{(2,0)}(X)$ and $H^{(0,2)}(X)$, the latter one being positive definite. Denoting $g:=\mathrm{dim}\,H^{(2,0)}(X)$, we can decompose $n_+$ as $n_+=2g+\tilde{n}_+$, where $\tilde{n}_+$ is the number of positive eigenvalues of the intersection form restricted to $H^{(1,1)}(X)$. Then the Hodge index theorem states that $\tilde{n}_+=1$, i.e. the intersection form on $H^2(X,\mathbb{R})$ has signature $(2g+1,n_-)$. Many of the surfaces that we are working with in this paper are non-compact, therefore this result is not directly applicable, but it is useful to keep it in mind. ${\scriptstyle \blacksquare}$

\subsection{The Chern-Weil formulas}\label{ChernWeil}

The so-called Chern-Weil theory allows to calculate topological invariants (of manifolds and, more generally, of vector bundles over these manifolds) in a differential-geometric way. Here we will be interested in expressions for the Euler characteristic and signature of the surfaces under consideration in terms of the K\"ahler metric on these surfaces. Such expressions can be found, for instance, in \cite{EGH}, and they are reviewed below as well. 

As before, we will assume that $X$ is a 4-manifold, possibly with boundary $\dd X$, although for the moment we will not require that it is K\"ahler. We will denote the Euler characteristic of a  by $\mathrm{Eu}(X, \dd X)$ and the signature by $\mathrm{Sgn}(X, \dd X)$. The simplest case is when $X$ does not have a boundary, i.e. $\dd X= \emptyset$, --- then $\mathrm{Eu}(X, \emptyset)$ and $\mathrm{Sgn}(X, \emptyset)$ are expressed through the curvature tensor as follows\footnote{Note that the factor of $\sqrt{g}$ is to appear in the denominator for the integral to be reparametrization-invariant. The expression acquires a more canonical form if expressed in the local frame, in terms of the (inverse) vierbein $E_A^a$, $R_{AB}:=E_A^a\,E_B^b\,R_{abmn}\,dx^m \wedge dx^n$. In that case $\mathrm{Eu}(X, \emptyset)=\frac{1}{32 \pi^2}\;\int\;d^4x\; \epsilon^{ABCD}\,R_{AB}\wedge R_{CD}$.}:
\bear
&&\mathrm{Eu}(X, \emptyset)=\frac{1}{32 \pi^2}\;\int\;d^4x\;\frac{1}{\sqrt{g}} \;\epsilon^{abcd}\,\epsilon^{mnpq}\,R_{abmn} \,R_{cdpq}\\
&&\mathrm{Sgn}(X, \emptyset)=
-\frac{1}{24 \pi^2}\;\int\,d^4x \;\epsilon^{abcd}\;R^{m}_{\;n ab}\, R^n_{\;m cd}
\eear
Note that on a K\"ahler manifold the only nonzero components of the curvature tensor are $R^a_{\,b}:=R^a_{\,bm\bar{n}}\,dz^m\wedge d\bar{z}^n=\frac{\dd \Gamma^{a}_{bm}}{\dd \bar{z}^n}$  (here the unbarred indices are holomorphic, and the barred indices are antiholomorphic) and its complex conjugate. Using this, the above integral for the Euler number may be simplified\footnote{In the formulas (\ref{inteu}) and (\ref{intsgn}) $R$ is the holomorphic part of the curvature tensor.}:
\bea\label{inteu}
\mathrm{Eu}(X, \emptyset)=
\frac{1}{8\pi^2}\;\int\,d^4x \,\big(\tr(R\wedge R)-\tr(R)\wedge\tr(R)\big)
\eea
Analogously one can write an expression for the signature:
\bea\label{intsgn}
\mathrm{Sgn}(X, \emptyset)=
-\frac{1}{24 \pi^2}\;\int\,d^4x \;\tr(R\wedge R+\bar{R} \wedge \bar{R})
\eea
In this paper we are restricting to the situation when the surface has $U(2)$ isometry group --- in this case both integrands depend only on $x=|z_1|^2+|z_2|^2$, therefore the integrals can be performed explicitly in terms of the K\"ahler potential $K(x)$. In fact, a much more economical way to write the corresponding formulas is to introduce a new function $Q(x):=x K'(x)$, in terms of which the topological invariants of $X$ acquire the following simple form:
\bear\label{eudens}
&&\!\!\!\!\!\!\mathrm{Eu}(X, \emptyset)=\left( 
-\frac{x^2 Q'(x)^2}{Q(x)^2}+\frac{2 x \left(x
   Q''(x)+Q'(x)\right)}{Q(x)}-\frac{2 x Q''(x)}{Q'(x)}\right)\vline_{\,0}^{\,\infty}
\\
\label{sgndens}
&& \!\!\!\!\!\!\mathrm{Sgn}(X, \emptyset)=\frac{1}{3} \left(\frac{3 x^2 Q'(x)^2}{Q(x)^2}+\frac{x^2
   Q''(x)^2}{Q'(x)^2}-\frac{2 x \left(x Q''(x)+3
   Q'(x)\right)}{Q(x)}+\frac{2 x Q''(x)}{Q'(x)} 
   \right)\vline_{\,0}^{\,\infty}
\eear
The point $x=0$ corresponds to the blow-up, so we expect that it is the contribution at zero that characterizes the blow-up topology.

In the formulas above we have assumed that the manifold $X$ has no boundary. For $\dP_1$ this is indeed the case. Using its K\"ahler potential, which is given by (\ref{kahdp1}) and (\ref{quoteqs1}), in the formulas (\ref{eudens}) and (\ref{sgndens}) above, we obtain the correct values
\bea
\mathrm{Eu}(\dP_1)=4,\quad\quad\mathrm{Sgn}(\dP_1)=0
\eea
The Euler characteristic is greater by one than that of $\CP^2$, since $\dP_1$ has an additional 2-cycle (the glued-in copy of $\CP^1$). The signature is zero, since, apart from the positive self-intersection cycle of $\CP^2$ (which is given by the hyperplane section), $\dP_1$ has one negative self-intersection cycle --- the exceptional divisor of the blow-up. We expect that a similar analysis should give correct results for the topology of all $\dP_{n}$ surfaces, however for larger $n$ technically the task is more complicated due to the smaller symmetry (isometry) group, and we leave it for future investigation.

However, in many cases, including the ones which will be of interest to us in subsequent sections, the manifold $X$ has  a boundary. If $\dd X\neq \emptyset$, there are additional contributions to the Euler characteristic and signature. First of all, both the Euler characteristic and signature receive a contribution that describes how the boundary is `embedded' in $\bar{X}$ (the closure of $X$). We will call such contributions $\delta \,\mathrm{Eu}[\dd X\subset \bar{X}]$ and $\delta \,\mathrm{Sgn}[\dd X\subset \bar{X}]$. They depend on the second fundamental form $\mathrm{II}(\dd X\subset \bar{X})$ and, in particular, they vanish if $\dd X$ is a totally geodesic submanifold of $X$. In fact, the boundary correction may be derived solely from the requirement that it has to vanish for a metric that is a  `product metric' near the boundary \cite{EGH} (i.e. when the boundary has zero second fundamental form). This is reviewed in Appendix \ref{ChernSimons} and leads to the following result:
\bear\label{delta1}
&&\delta \,\mathrm{Eu}[\dd X\subset \bar{X}]=\frac{-1}{32\pi^2}\,\int\limits_{\dd X}\,2\,\epsilon^{abcd} \big( \zeta_{ab} R_{cd}-{2\over 3} \zeta_{ab} (\zeta^2)_{cd}\big)\,d^3 x\\ \label{delta2}
&&\delta \,\mathrm{Sgn}[\dd X\subset \bar{X}]=\frac{1}{24\pi^2}\,\int\limits_{\dd X}\,\tr(\zeta\wedge R),
\eear
where $\zeta$ is the second fundamental form. If we assume that the boundary is defined by the equation $E_0=0$, where $E_0$ is a component of the tetrad, $\zeta$ can be expressed through the connection $\omega$ as $\zeta_{0i}=\omega_{0i}=-\zeta_{i0}$, all other components being zero. 

For a metric coming from a $U(2)$-invariant K\"ahler potential these formulas give the following result:
\bear
&&\delta \,\mathrm{Eu}[\dd X\subset \bar{X}]=
\frac{x^2 Q'(x)^2}{Q(x)^2}-\frac{2 x \left(x Q''(x)+Q'(x)\right)}{Q(x)}+\frac{2 x Q''(x)}{Q'(x)}+2 |_{x\to x_{\textrm{boundary}}}
\\
&&\delta \,\mathrm{Sgn}[\dd X\subset \bar{X}]=-\frac{\left(x Q'(x)^2-Q(x) \left(x
   Q''(x)+Q'(x)\right)\right)^2}{Q(x)^2 Q'(x)^2}\big|_{x\to x_{\textrm{boundary}}}
\eear

There is yet another contribution to the signature, called $\eta[\dd X]$, which depends on the intrinsic properties of the boundary manifold $\dd X$ only \cite{APS1}. We will not attempt to explain the origin of $\eta[\dd X]$ and we refer the interested reader to the original work \cite{APS1,APS2} or to the review \cite{EGH}. The complete formulas for the topological invariants look as follows:
\bear\label{CW1}
&&\mathrm{Eu}(X, \dd X)=\mathrm{Eu}(X, \emptyset)+\delta \,\mathrm{Eu}[\dd X\subset \bar{X}]\\ \label{CW2}
&&\mathrm{Sgn}(X, \dd X)=\mathrm{Sgn}(X, \emptyset)+\delta \,\mathrm{Sgn}[\dd X\subset \bar{X}]+\eta[\dd X]
\eear
The calculation of $\eta[M]$ has to be done separately for every manifold $M$, but the only case that will be of interest to us is that of the lens space $M=L(m,1):=S^3/\mathbb{Z}_m$. If the sphere $S^3$ is defined by the equation $|w_1|^2+|w_2|^2=1$, then the action of $\mathbb{Z}_m$ is given by $(w_1, w_2)\to (\omega \,w_1, \omega \,w_2)$, where $\omega$ is the $m$-th root of unity, $\omega=e^{{2\pi i \over m}}$. In this case the $\eta$-invariant is \cite{EGH}:
\bea\label{lens}
\eta[L(m,1)]=\frac{1}{m}\sum\limits_{k=1}^{m-1} \;\left(\mathrm{ctg}\frac{\pi k}{m}\right)^2=\frac{1}{3m}\left(m^2-3m+2\right)
\eea
We will use the above formulas in the end of the next Section to calculate the Euler characteristic and signature of the K\"ahler-Einstein spaces $Y_m$ described below.

\section{K\"ahler-Einstein metrics}\label{KEmetr}

We wish to consider the K\"ahler metric $g_{i\bar{j}}$ arising from an  $U(2)$-invariant potential with $K=K(|z_1|^2+|z_2|^2)=K(x)$. Such a metric is
\bea\label{metric1}
g_{i\bar{j}}=\frac{\dd^2 K}{\dd z^i \dd \bar{z}^j}=\delta_{ij}\,K'+z_j \bar{z}_i K''
\eea
To find out in which case this metric is positive-definite we calculate the norm $\| v\|$ of an arbitrary vector $v$:
\bea
\|v\|=\bar{v_i} v_i \, K'+|\bar{v}_i z_i|^2 K''
\eea
If $K'>0$ and $K''>0$, this is obviously positive. If instead $K'>0$ but $K''<0$, then we use the Schwartz inequality to obtain
\bea
\| v\| \geq \bar{v_i} v_i \, K'+\bar{v_i} v_i \, x\,K''=\bar{v_i} v_i\,(xK')'
\eea
Note that the Schwartz inequality is saturated for $v\sim z$, so this is a sharp estimate. Hence the necessary and sufficient condition for the metric to be positive-definite is that
\bea\label{posdef}
(xK')'>0
\eea
The line element in real coordinates for the metric (\ref{metric1}) is written out in Appendix \ref{metric}. The explicit expression (\ref{lineel}) confirms our conclusion that the positivity of the metric requires $K'>0, (xK')'>0$.

The Einstein condition
\bea\label{Einstein}
R_{i\bar{j}}=a \,g_{i\bar{j}}
\eea
for a $U(2)$-invariant K\"ahler metric $g_{i\bar{j}}=\frac{\dd^2 K}{\dd z^i \dd \bar{z}^j}$ with $K=K(x)$ leads to the equation
\bea
\frac{d}{dx}\big(\log \det g_{i\bar{j}}\big) = -a K'
\eea
Introducing a new function $Q:=x K'$, we can recast this equation in the form
\bea\label{maineq}
\frac{d}{dx} \big( x(Q^2)'-2Q^2+a\,{2\over 3} Q^3\big)=0
\eea
Everywhere below (except for the last \emph{Comment} in this Section) we will be dealing with manifolds of negative curvature, hence we will set $a=-1$. Explicit integration gives ($b$ is a constant of integration):
\bear\label{eqy}
&&\log{x}=\sum\limits_{i=1}^3\; \frac{\log{(Q-y_i)}}{2+y_i},\quad\textrm{where}\quad y_i: \;y_i^3+3 y_i^2+b=0\\ \label{Qx}
&& \label{Q}\Rightarrow x=\prod\limits_{i=1}^3\;(Q-y_i)^{1\over 2+y_i}
\eear
The positivity condition (\ref{posdef}) requires that $Q$ is a growing function of $x$. It is easy to check explicitly that it is satisfied for $Q>\underset{i}{\mathrm{max}} \,y_i:=y$. It follows from (\ref{Qx}) that the behaviour of $Q$ in the vicinity of $y$ ($Q\to y$ implies $x\to 0$) is $Q=y+x^{2+y}+\ldots$. Recall that $Q=x K'$, so this behavior for $Q$ implies that the K\"ahler potential has the folowing form close to $x=0$:
\bea\label{expnear0}
K=y \log{x}+{1\over 2+y} x^{2+y}+\ldots,
\eea
which is clearly the same as (\ref{Kahlerexpansion}) with $m=2+y$. The fact that $y=m-2$ should be a solution to $y^3+3 y^2+b=0$ implies that $b=-(m-2)^2(m+1)$.

\vspace{0.3cm}
\emph{Comment.} Note that the expansion (\ref{expnear0}) is only valid for $y>0$. If we set $b=0$ in (\ref{eqy}), (\ref{Qx}), the solution will describe pure Lobachevsky space, and the expansion at $x\to 0$ is $Q\sim3x+3x^2+\ldots$ Clearly, one cannot obtain the Eguchi-Hanson solution from (\ref{Qx}), since (\ref{Qx}) corresponds to $a=-1$ (negative curvature), whereas the Eguchi-Hanson space has $a=0$. $\footnotesize \blacksquare$

\vspace{0.3cm}
Let us find out how $x$ behaves at large $Q$. It is a simple property of the cubic equation of the form (\ref{eqy}) that the following fact holds: $\sum\limits_{i=1}^3\frac{1}{2+y_i}=0$. It then follows from (\ref{eqy}) that for $Q\to\infty: x\to 1$, i.e. the total space of the line bundle $\mathcal{O}(-m)$ for $m\geq 3$ is a ball $|z_1|^2+|z_2|^2<1$! To learn how the metric behaves near the boundary of the ball we expand (\ref{Qx}): $x=1-\frac{3}{Q}+\ldots$, which leads to $Q\sim \frac{3}{1-x}$ for $x\to 1$. For the K\"ahler potential this implies, in the vicinity of the boundary $x=1$:
\bea\label{kahbound}
K=-3\log{(1-|z_1|^2-|z_2|^2)}+\ldots\quad \textrm{for}\quad |z_1|^2+|z_2|^2\to 1
\eea
This is in fact nothing but the K\"ahler potential on the four-dimensional Lobachevsky space $H_4$! Indeed, the Lobachevsky space $H_4$ may be thought of as the quotient
\bea
H_4=\frac{U(1,2)}{U(1)\times U(2)}
\eea
and the K\"ahler potential in homogeneous coordinates on such a space is
\bea
K_{\mathrm{H}_4}\sim -\log{(|z_0|^2-|z_1|^2-|z_2|^2)},
\eea
which translates into (\ref{kahbound}) once we choose the `gauge' $z_0=1$ and use $z_1, z_2$ as inhomogeneous coordinates.

\begin{figure}
\centering
\includegraphics[width=0.8\textwidth]{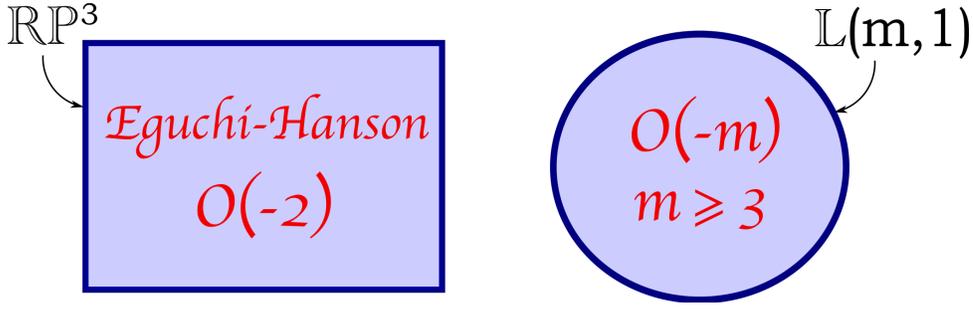}
\caption{Total spaces of the line bundles $\mathcal{O}(-m)$ for different $m$. Here $\mathds{L}(m,1)$ is the lens space, and $\mathds{RP}^3=\mathds{L}(2,1)$.}
\label{fig:awesome_image}
\end{figure}

Our claim is that the metric described by the K\"ahler potential, whose $Q$-function (related to $K$ by $Q=xK'$) is given by equation (\ref{Q}) with $b=-(m-2)^2(m+1),\;\;m=3, 4, \ldots$, is the metric on
\bea
Y_m:=\textrm{the total space of the line bundle} \; \mathcal{O}(-m) \; \textrm{over} \;\CP^1 \;
\eea
To prove it we need to consider the topological data. First of all, since $Y_m$ is the total space of a bundle with a contractible fiber, its cohomologies are the same as those of $\CP^1$. In particular, $\mathrm{Eu}(Y_m)=2$. Since $Y_m$ is the total space of a negative bundle over $\CP^1$, the general logic of Section \ref{selfintsec} tells us that $\mathrm{Sgn}(Y_m)=-1$ (since the self-intersection of the glued-in $\CP^1$ is the degree of the normal bundle, it follows that $n_-=1$).\vspace{0.3cm}

\emph{Comment.} The two other related cases are those of positive and zero normal bundle, i.e. respectively of, say, $\CP^2$ and of $\CP^1\times \CP^1$. In the case of $\CP^2$ we can use the formula (\ref{sgndens}) with $Q=x{d\over dx}\log{(1+x)}={x\over 1+x}$ to obtain $\mathrm{sgn}(\CP^2)=1$. In the case of $\CP^1\times \CP^1$ the matrix $R_{ab}$ representing the Riemann tensor splits:
\bea
R:=\{R_{ab}\}=\left(\scalemath{1}{
\begin{array}{cc}
r_1\, dz_1\wedge d\bar{z}_1&0\\0&r_2\, dz_2 \wedge d\bar{z}_2
\end{array}}\right),
\eea
for some functions $r_{1,2}$, therefore $\tr R^2=0$, so $\mathrm{Sgn}(\CP^1\times \CP^1)=0$. This is compatible with the fact that two lines representing two distinct homology classes (the `left' and `right' factors of $\CP^1\times \CP^1$) are embedded with trivial normal bundle and intersect at one point (see Fig. \ref{linesfig}). The intersection matrix is $\left(\begin{array}{cc}0&1\\1&0\end{array}\right)$ and has eigenvalues $\pm 1$. ${\scriptstyle \blacksquare}$
\vspace{0.3cm}

Using the Chern-Weil formulas (\ref{CW1}), (\ref{CW2}), we can now check that the Euler characteristic and signature of the manifolds $Y_m$ are precisely as required. It turns out that in the bulk terms in (\ref{CW1}), (\ref{CW2}) the only $m$-dependent contribution comes from the point $x=0$, i.e. from the blown-up sphere. If we assume that the boundary is a sphere $S^3$, for which $\eta[S^3]=0$, a direct calculation gives the following answers:
\bear\label{eu1}
&&\mathrm{Eu}=2m \\ \label{sgn1}
&&\mathrm{Sgn}=-{1\over 3}\,(2+m^2)
\eear
This is not the answer we need. However, we claim that in order to get the right answer we should take a further quotient
\bea\label{quotnew}
(z_1, z_2)\sim e^{{2\pi i\over m}}\,(z_1, z_2)
\eea
Clearly, in the integral terms in the formulas (\ref{CW1}), (\ref{CW2}) this amounts to restricting the integration range over one of the angles $\phi$ to $[0,{2\pi \over m})$ instead of $[0, 2\pi)$. Hence the values in the right hand sides of (\ref{eu1}), (\ref{sgn1}) are simply divided by $m$. However, the quotient (\ref{quotnew}) alters the boundary of our 4-manifold --- indeed, it is no longer $S^3$ but rather the lens space $\mathds{L}(m, 1)=S^3/\mathds{Z}_m$ (see the definition above, before (\ref{lens})). And the $\eta$-invariant for the lens space is no longer zero! Instead, it is given by (\ref{lens}), which enters as an additional contribution to the signature. As a result, we obtain the final answers for the spaces $Y_m$ with the quotient (\ref{quotnew}) taken into account:
\bear\label{eu2}
&&\mathrm{Eu}(Y_m)=2m\cdot{1\over m}=2 \\ \label{sgn2}
&&\mathrm{Sgn}(Y_m)=-{1\over 3}\,(2+m^2)\cdot{1\over m}+\frac{1}{3m}\left(m^2-3m+2\right)=-1
\eear

Note that the analogous calculation for the Eguchi-Hanson space leads to the same result as above, but in that case the $\eta$-term is zero, since the boundary is $\mathds{RP}^3=\mathds{L}(2, 1)$ and $\eta[\mathds{RP}^3]=0$.

Summarizing, it turns out that the contribution of the point $x=0$ coming from the `singular' behavior of the K\"ahler potential (\ref{Kahlerexpansion}), i.e. the contribution of the glued in $\CP^1$, exactly cancels the $\eta$-contribution of the boundary lens space $\mathds{L}_m$! Hence we have a very nontrivial cancellation of the terms coming from the boundary of $Y_m$ versus the terms from the deep interior (the `horizon').

Another fact about the manifolds $Y_m$ that is worth checking is that their first Chern number is equal to $2-m$. Indeed, we expect that\footnote{See the \emph{Comment} after formula (\ref{holform}).} $c_1(Y_m)=(2-m)\, \omega_\mathds{FS}$, where $\omega_\mathds{FS}$ is the Fubini-Study form, or any form that has period $1$ when integrated over the blown-up sphere $\mathcal{S}$ (which is the only homologically nontrivial sphere in $Y_m$). Therefore
\bea
\int\limits_{\mathcal{S}} \,c_1(Y_m)=2-m
\eea
To check this formula we make a change of variables $z=z_1, w=\frac{z_2}{z_1}$. The submanifold $\mathcal{S}=\{z\to 0, w\; \textrm{arbitrary}\}$ is the relevant $\CP^1\subset Y_m$. In the limit $z\to 0$ the integral
$ \frac{i}{2\pi}\,\int\limits_\mathcal{S} R_{i\bar{j}}\,dz_i\wedge d\bar{z}_j$ may be calculated, using $R_{i\bar{j}}=-g_{i\bar{j}}$, from the first term in the expansion of the K\"ahler potential (\ref{expnear0}), i.e. from the K\"ahler potential $K=(m-2)\,\log{(1+|w|^2)}$ on $\mathcal{S}=\CP^1$:
\bea
\int\limits_{\mathcal{S}} \, c_1(Y_m)=-(m-2)\,\frac{i}{2\pi}\,\int\limits_\mathcal{S} \dd_w \dd_{\bar{w}}  \log{(1+|w|^2)} \,dw\wedge d\bar{w}=2-m
\eea

\emph{Comment.} It is interesting to note that the Chern-Simons type contribution from the boundary in the case of $Y_m$ is equal to ${1\over m}\times$ (the corresponding value for Lobachevsky space). The ${1\over m}$ factor comes simply from the fact that the range of integration over the angular variables is $m$ times smaller than that of an $S^3$. The fact that the Chern-Simons term does not give a more nontrivial contribution to the signature stems from the fact that the expansion of $Q$ near the boundary, i.e. around $x\to 1$, is as follows:
\bea
Q(Y_m)=\frac{3}{1-x}-3+a_m \,(1-x)^2+\ldots
\eea
The important fact to observe is that the linear term $O(1-x)$ is missing. If the expansion of $Q$ involved a linear term, i.e. $Q(Y)=\frac{3}{1-x}-3+\alpha\,(1-x)+\ldots$, there would be yet another contribution to the signature:
\bea
\delta \mathrm{Eu}=0 ,\quad\quad \delta \mathrm{Sgn}=-{8\over 3}\,\alpha
\eea
Interestingly, $\mathrm{eu}(X, \emptyset)\big|_x+\delta \mathrm{eu}(\dd X \subset \bar{X})=2$ is a constant --- the total contribution of the boundary of the ball to the Euler characteristic does not depend on $Q(x)$ at all, which is the reason why the behavior of $Q$ near the boundary does not affect $\mathrm{Eu}$. ${\scriptstyle \blacksquare}$

\vspace{0.3cm}
\emph{Comment.} A theorem of Matsushima \cite{Besse} states that, if the automorphism group of a complex manifold is not reductive, then it cannot admit a K\"ahler-Einstein metric. In particular, this is the case for the del Pezzo surface of rank $1$ that we considered above. Using the ideas developed here, we can establish the non-existence of a geodesically complete K\"ahler-Einstein metric on the tubular neighborhood of the blow-up (i.e. on the total space of the line bundle $\mathcal{O}(-1)$). The total space of $\mathcal{O}(-1)$ has first Chern number $c_1=1$, therefore we expect to be dealing with a manifold of positive curvature. Therefore we set $a=1$ in (\ref{Einstein}). This leads to the solution of (\ref{maineq}):
\bea
x=\prod\limits_{i=1}^3\;|Q+y_i|^{1\over 2+y_i},\quad\textrm{where}\quad y_i:\;y_i^3+3 y_i^2-2=0
\eea
The solution is well-defined (meaning that $Q>0$ and $Q'>0$) for $x\in[0,\infty)$, and the function $Q$ takes the respective values $[1,1+\sqrt{3})$. Such behavior is in fact similar to the one of the (positively-curved) $\CP^2$, in which case $Q_{\CP^2}=\frac{x}{1+x}$ and $Q\to 1$ as $x\to \infty$. What is different in these two cases is the (subleading) behavior of $Q$ at $x\to \infty$, namely for $\CP^2$: $Q=1-{1\over x}+\ldots$, whereas for the solution under consideration $Q=(1+\sqrt{3})-2 \sqrt{3}\,(2x)^{1-\sqrt{3}}+\ldots$ If one then uses the bulk formulas (\ref{eudens}), (\ref{sgndens}) for $\mathrm{Eu}$ and $\mathrm{Sgn}$, one obtains irrational answers for these topological characteristics, which implies that the corresponding space is not a manifold.
${\scriptstyle \blacksquare}$

\section{Conclusion}

In this paper we provided evidence that in the $U(2)$-invariant case the normal bundle to the holomorphic sphere $\CP^1 \subset X$ in a complex surface $X$ is captured by the characteristic behavior of the K\"ahler potential in the vicinity of this line. More exactly, the following formula holds for a line with normal bundle $\mathcal{O}(-m),\;m\geq 1$:
\begin{empheq}[box=\fbox]{align}
\label{evopxxx}\nonumber
\renewcommand{\arraystretch}{2.5}
\begin{tabular}{m{0.8 \textwidth}}
\vspace{-0.6cm}
$\quad K(x)=a \log\,x+b\, x^m+ \ldots\;\;\quad\textrm{as}\;\;x=|z_1|^2+|z_2|^2\to 0,\quad a, b > 0 $
\end{tabular}
\hspace{1em}
\end{empheq}
A natural question which arises is whether, knowing the characteristic behavior of the K\"ahler potential at $x\to 0$, we can reconstruct the K\"ahler potentials on the total spaces of the line bundles $\mathcal{O}(-m)$ over $\CP^1$ with some desirable geometric properties. We have addressed this question, requiring that the K\"ahler metric satisfies Einstein's equations. It turns out that the corresponding equations have a solution of negative cosmological constant (negative curvature) for $m\geq 3$. This solution looks as follows:
\begin{empheq}[box=\fbox]{align}
\label{evopxxx}\nonumber
\hspace{1em}
Y_m: \quad Q=xK',\quad x=\prod\limits_{i=1}^3\;(Q-y_i)^{1\over 2+y_i},\;\;\textrm{where}\;\; y_i^3+3 y_i^2-(m-2)^2(m+1)=0
\hspace{1em}
\end{empheq}

The interesting fact is that, for $m \geq 3$, the K\"ahler potential tends to infinity as $|z_1|^2+|z_2|^2\to 1$. Moreover, asymptotically near $|z_1|^2+|z_2|^2 \simeq 1$ the metric behaves as the Lobachevsky space $\mathrm{H}_4$ metric near the boundary. However, the requirement that the topological characteristics of this space --- the Euler characteristic and signature ---  are integers, implies that the boundary cannot be $S^3=\dd \mathrm{H}_4$, but it rather has to be a quotient thereof, more precisely the lens space $L(m,1)=S^3/\mathbb{Z}_m$. Requiring that the boundary is the appropriate lens space, we find that there is a nontrivial cancellation between the contributions to the topological numbers from the boundary and the $\CP^1$ glued in at the origin (the `horizon'), which is just right to provide $Y_m$ with the topological numbers of a line bundle over $\CP^1$. Therefore, the spaces described by these metrics can be called `asymptotically locally hyperbolic', in analogy with the well-known ALE spaces.

Interestingly, there appears to be yet another set of metrics on the total space of $\mathcal{O}(-m)$ for $m\geq 3$, discovered by LeBrun \cite{lebrun}. These metrics are K\"ahler and possess an anti-self-dual Weyl tensor. Their Ricci scalar $R$  is zero, and they are asymptotically-locally-Euclidean, rather than hyperbolic.

\begin{center}
\line(1,0){150}
\end{center}
\vspace{-0.6cm}
\begin{center}
{\normalfont\scshape \large Acknowledgments}
\end{center}
\vspace{-0.9cm}
\begin{center}
\line(1,0){150}
\end{center}

I would like to thank I.Ya. Aref'eva, P. Di Vecchia, S. Gorchinskiy, V. Przhiyalkovskiy, K. Zarembo for interesting discussions. I am indebted to Prof. A.A.Slavnov and to my parents for constant support and encouragement. My work was supported in part by grants RFBR 11-01-00296-a, 13-01-12405 ofi-m2, 12-01-31298-mol-a and in part by grant for the Support of Leading Scientific Schools of Russia NSh-4612.2012.1.
\newpage
\begin{center}
\line(1,0){150}
\end{center}
\appendix
\vspace{-0.6cm}
\begin{center}
{\normalfont\scshape \large Appendix}
\end{center}
\vspace{-0.9cm}
\begin{center}
\line(1,0){150}
\end{center}

\section{Relation of K\"ahler potential to the moment map}\label{Kahmomrel}

In the case that the symplectic manifold is K\"ahler there is an interesting and simple relation of the moment map (for the action of a group $G$) to the K\"ahler potential. We assume that the K\"ahler potential is invariant under $G$ (see the \emph{Comment }at the end of this Section). To obtain this relation we first of all recall the well-known expression for the symplectic form
\bea\label{omega}
\omega=i\dd\bar{\dd}K=i\dd_i\bar{\dd}_jK\,dz^i\wedge d\bar{z}^j
\eea
The usual relation of the moment map to the symplectic form is
\bea\label{mommap}
d\mu(a)=\omega(\bullet ,V_a),
\eea
where $V_a$ is a vector field corresponding to the Lie algebra element $a$. Decomposing the vector field into holomorphic and antiholomorphic parts, we obtain:
\bea\label{vector}
V_a=v_a+\bar{v}_a
\eea
Here we mean that $v_a$ is a holomorphic vector field, i.e. $v_a=\sum\,v_a^i\,{\dd\over \dd z^i}$ with the \emph{coefficients $v_a^i$ being holomorphic} as well. In other words, $\bar{\dd}_j\,v_a^i=0$. For what follows we introduce the complex structure operator $\cst$, which acts on $v$ as
\bea
\cst V_a=i\big( v_a-\bar{v}_a\big)
\eea
Then the formula (\ref{mommap}) for the K\"ahler form (\ref{omega}) and vector field (\ref{vector}) decomposes into the $dz$ and $d\bar{z}$  parts:
\bea\label{holmom}
\dd_i\mu=i \,\dd_i\bar{\dd}_jK\,\bar{v}^j,\quad \bar{\dd}_j \mu=-i \,\dd_i\bar{\dd}_jK\, v^i
\eea
Since the $v^i$ components of $v$ are holomorphic and $\bar{v}^j$ are antiholomorphic, the above equations can be rewritten as
\bea
\dd_i \big( \mu-i \,\bar{\dd}_jK\,\bar{v}^j \big) =0 ,\quad \bar{\dd}_j \big( \mu+i \,\dd_i K\, v^i \big) = 0
\eea
Solving the first equation, we get
\bea\label{mom2}
\mu=i \,\bar{\dd}_jK\,\bar{v}^j+\bar{f}(\bar{z})
\eea
$\mu$ is a real function, which implies
\bea\label{k3}
i \,\bar{\dd}_jK\,\bar{v}^j+\bar{f}(\bar{z})=-i \,\dd_jK\,v^j+f(z)
\eea
Since we have assumed that $K$ is invariant under $v$,
\bea\label{kah2}
i \,\dd_jK\,v^j+i \,\bar{\dd}_jK\,\bar{v}^j=0,
\eea
hence $f(\bar{z})=\bar{f}(z)$, which means that $f$ is a real constant: $f=c$. Hence, using (\ref{mom2}) and (\ref{kah2}), we can write $\mu$ as
\bea
\mu={1\over 2} \nabla_{\cst v} K+c,
\eea
which is the desired relation.\newline

\vspace{0.3cm}
\emph{Comment.} Note that, if $K$ is not $G$-invariant, we can define $K=K_1+k(z)+\bar{k}(\bar{z})$, where $k(z)$ satisfies $i \,\dd_j k\,v^j=f(z)$. In this case (\ref{k3}) implies that $K_1$ is $G$-invariant. The addition of the holomorphic function $k(z)$ to the K\"ahler potential is inessential, since it does not affect the K\"ahler metric --- it is the usual redundancy in the definition of the K\"ahler potential. Hence we can assume, without loss of generality, that the K\"ahler potential is $G$-invariant. ${\scriptstyle \blacksquare}$

\section{Derivation of the K\"ahler quotient formula}\label{appKahquot}

In this Appendix we provide a direct proof of the formula (\ref{Kquot}) for the K\"ahler potential on the quotient, which is central for the derivation of the results presented in this paper. In other words, we need to prove that
\bea\label{pullbkahpot}
\omega:=i\,\dd_w\,\bar{\dd}_{\bar{w}}\;\pi^\ast(K)=\pi^{\ast}(\tilde{\omega}\big|_{\mu^{-1}(0)})
\eea

Let us compute directly the r.h.s. To do it we substitute $z=z(w, \bar{w})$ into the expression $\tilde{\omega}=i\,\dd_i\bar{\dd}_jK\,dz^i\wedge d\bar{z}^j$ for the symplectic form\footnote{Everywhere below $\dd_i\bar{\dd}_j \tilde{K}\,$ denotes the derivatives w.r.t. $z^i, \bar{z}^j$: $\dd_i\bar{\dd}_j \tilde{K}:=\frac{\dd^2 \tilde{K}}{\dd z_i\,\dd\bar{z}_j}$.} (recall that $z(w, \bar{w})$ is a point in $\mu^{-1}(0)\subset \mathcal{M}$, whereas $w$ is a point in the ambient space $\mathcal{M}$):
\bear \nonumber
\pi^{\ast}(\tilde{\omega}\big|_{\mu^{-1}(0)})&=& i\,\dd_i\bar{\dd}_j \tilde{K}\;\frac{\dd z_i}{\dd w_k}\;\frac{\dd \bar{z}_j}{\dd \bar{w}_s}\;dw_k\wedge d\bar{w}_s+i\,\dd_i\bar{\dd}_j\tilde{K}\;\frac{\dd z_i}{\dd \bar{w}_k}\;\frac{\dd \bar{z}_j}{\dd w_s}\;d\bar{w}_k\wedge d w_s+\\ 
&+& i\,\dd_i\bar{\dd}_j\tilde{K}\;\frac{\dd z_i}{\dd w_k}\;\frac{\dd \bar{z}_j}{\dd w_s}\;d w_k\wedge d w_s+i\,\dd_i\bar{\dd}_j\tilde{K}\;\frac{\dd z_i}{\dd \bar{w}_k}\;\frac{\dd \bar{z}_j}{\dd \bar{w}_s}\;d\bar{w}_k\wedge d \bar{w}_s
\eear

\vspace{0.3cm}
Since this ought to be a K\"ahler form in the $w, \bar{w}$ coordinates, it has to be of type $(1,1)$. For this to be the case, the terms of the form $d w_k\wedge d w_s,\;\; d\bar{w}_k\wedge d \bar{w}_s$ have to be zero. And, clearly, they cannot cancel each other, so they need to be zero separately. Let us therefore consider one of these terms, for example,
\bea
\Delta:=i\,\dd_i\bar{\dd}_j\tilde{K}\;\frac{\dd z_i}{\dd w_k}\;\frac{\dd \bar{z}_j}{\dd w_s}\;d w_k\wedge d w_s
\eea

The dependence of the $z,\; \bar{z}$ variables on $w, \; \bar{w}$ is not holomorphic, however the nonholomorphicity comes solely from the dependence of the $t$'s (the transformation parameters of $G_{\CC}$) on both $w, \;\bar{w}$. For example, $\bar{z}=\bar{z}(t,\;\bar{w})$, therefore
\bea
\frac{\dd \bar{z}_j}{\dd w_s}=\frac{\dd \bar{z}_j}{\dd t_a}\;\frac{\dd t_a}{\dd w_s}=-i\,\bar{v}_a^{j}\;\frac{\dd t_a}{\dd w_s} ,
\eea
where in the latter equality we have used (\ref{floweqs}). Plugging this into $\Delta$, we get
\bea
\Delta = \dd_i\bar{\dd}_j\tilde{K}\;\bar{v}_a^{j}\;\frac{\dd t_a}{\dd w_s}\;\frac{\dd z_i}{\dd w_k}\;\;d w_k\wedge d w_s
\eea

Using the formula (\ref{holmom}) for $\dd_i\mu$, we get
\bea
\Delta =  -i \,\frac{\dd \mu_a}{\dd z^i}\;\frac{\dd z_i}{\dd w_k}\;\frac{\dd t_a}{\dd w_s}\;\;d w_k\wedge d w_s
\eea
From the definition of the change of variables $z=z(w, \bar{w}),\;\;\bar{z}=\bar{z}(w, \bar{w})$ we have $\mu_a(z(w, \bar{w}),\;\bar{z}(w, \bar{w}))=0$ identically, so taking the derivative w.r.t. $w_k$, we obtain:
\bea
\frac{\dd \mu_a}{\dd z^i}\;\frac{\dd z_i}{\dd w_k}=-\frac{\dd \mu_a}{\dd \bar{z}^i}\;\frac{\dd \bar{z}_i}{\dd w_k}=i\, \bar{v}_b^i\,\frac{\dd \mu_a}{\dd \bar{z}^i}\;\frac{\dd t_b}{\dd w_k},
\eea
so that $\Delta$ becomes
\bea
\Delta = \bar{v}_b^i\,\frac{\dd \mu_a}{\dd \bar{z}^i}\;\frac{\dd t_b}{\dd w_k} \;\frac{\dd t_a}{\dd w_s}\;\;d w_k\wedge d w_s
\eea
The moment maps $\mu_a$ transform covariantly under the group $G$, namely,
\bea
v_a^i\,\dd_i \mu_b+\bar{v}_a^i\,\bar{\dd}_i \mu_b=f_{abc}\,\mu_c
\eea
(if $G$ is abelian, $f_{abc}=0$, so the $\mu_a$'s are simply invariant under $G$). If one restricts this equation to $\mu^{-1}(0)$ (which is the case under consideration), the r.h.s. is zero, which reflects the simple fact that $\mu^{-1}(0)$ is $G$-invariant. Therefore
\bea
\bar{v}_b^i\,\frac{\dd \mu_a}{\dd \bar{z}^i}\big|_{\mu^{-1}(0)}=-{1\over 2\, i} \, i\left( v_b^i\,\frac{\dd \mu_a}{\dd z^i}-\bar{v}_b^i\,\frac{\dd \mu_a}{\dd \bar{z}^i}\right)=-{1\over 2 \,i} \nabla_{\mathrm{J}V_b}\mu_a
\eea
Using (\ref{momkahpot}), we can rewrite $\Delta$ as
\bea
\Delta={i\over 8 }\;[\nabla_{\mathrm{J}V_b}, \nabla_{\mathrm{J}V_a}] \,\tilde{K}\;\cdot\;\frac{\dd t_b}{\dd w_k} \;\frac{\dd t_a}{\dd w_s}\;\;d w_k\wedge d w_s
\eea
It is easy to see that $[\nabla_{\mathrm{J}V_b}, \nabla_{\mathrm{J}V_a}]\, \tilde{K}=\nabla_{[V_b, V_a]}\,\tilde{K}=f_{abc} \nabla_{V_c} \, \tilde{K}=0$ due to the $G$-invariance of the K\"ahler potential. Therefore $\Delta=0$, so that indeed $\pi^{\ast}(\omega\big|_{\mu^{-1}(0)})$ is of type (1,1).

\vspace{0.3cm}
Hence we are left with the following expression for the pull-back under $\pi$ of the restricted symplectic form:
\bea\label{derkah}
\pi^{\ast}(\omega\big|_{\mu^{-1}(0)})=i\,\dd_i\bar{\dd}_j\tilde{K}\,\left( \frac{\dd z_i}{\dd w_k}\;\frac{\dd \bar{z}_j}{\dd \bar{w}_s}-\frac{\dd z_i}{\dd \bar{w}_s}\;\frac{\dd \bar{z}_j}{\dd w_k} \right) \;dw_k\wedge d\bar{w}_s
\eea
We are now going to simplify this expression and demonstrate that it can be written as a double external derivative $i\,\dd_w\,\bar{\dd}_{\bar{w}}\;\pi^\ast(K)$, as stated in (\ref{pullbkahpot}). According to (\ref{Kquot}), $\pi^\ast(K)=\tilde{K}(z(w, \tau),\bar{z}(\bar{w},\tau))+2\,\sum\limits_{a=1}^n\,r_a\,\tau_a(w, \bar{w})$, so let us first of all calculate the derivative of $\tilde{K}(z(w, \tau),\bar{z}(\bar{w},\tau))$:
\bea
\frac{\dd}{\dd w_i} \tilde{K}(z(w, \tau),\bar{z}(\bar{w},\tau))= \frac{\dd}{\dd z_k} \tilde{K}(z(w, \tau),\bar{z}(\bar{w},\tau))\;\frac{\dd z_k}{\dd w_i}\big|_\tau+\frac{\dd}{\dd \tau_a} \tilde{K}(z(w, \tau),\bar{z}(\bar{w},\tau))\;\frac{\dd \tau_a}{\dd w_i},
\eea
where $\frac{\dd z_j}{\dd w_i}\big|_\tau$ means that we are taking a partial derivative with $\tau=\textrm{const.}$ Clearly
\bear
\frac{\dd}{\dd \tau_a} \tilde{K}(z(w, \tau),\bar{z}(\bar{w},\tau))&=&\frac{\dd}{\dd z_i} \tilde{K}\,\cdot \,\frac{\dd z_i}{\dd \tau_a}+\frac{\dd}{\dd \bar{z}_i} \tilde{K}\,\cdot \,\frac{\dd \bar{z}_i}{\dd \tau_a}=\\&=&i v^i_a\,\frac{\dd}{\dd z_i} \tilde{K}-i \bar{v}^i_a\,\frac{\dd}{\dd \bar{z}_i} \tilde{K}=\nabla_{\mathrm{J} V_a}\,\tilde{K}=2(\mu_a-r_a),
\eear
where the last equality follows from (\ref{momkahpot}). The moment map $\mu_a$ vanishes identically: $\mu_a(z(w, \tau), \bar{z}(\bar{w}, \tau))=0$. Hence
\bea
\frac{\dd}{\dd w_i} \tilde{K}(z(w, \tau),\bar{z}(\bar{w},\tau))= \frac{\dd}{\dd z_j} \tilde{K}(z(w, \tau),\bar{z}(\bar{w},\tau))\;\frac{\dd z_j}{\dd w_i}\big|_\tau-2 r_a\;\frac{\dd \tau_a}{\dd w_i}
\eea
Differentiating this equation w.r.t. $\bar{w}_j$ we obtain:
\bea\label{form1}
\frac{\dd^2 \tilde{K}}{\dd w_i \dd \bar{w}_j} =\frac{\dd^2 \tilde{K}}{\dd z_k \dd \bar{z}_m} \;\frac{\dd \bar{z}_m}{\dd \bar{w}_j}\big|_\tau\;\frac{\dd z_k}{\dd w_i}\big|_\tau+ \frac{\dd}{\dd \tau_b}\frac{\dd \tilde{K}}{\dd z_k}\;\frac{\dd \tau_b}{\dd \bar{w}_j}\;\frac{\dd z_k}{\dd w_i}\big|_\tau+\frac{\dd \tilde{K}}{\dd z_k} \;\frac{\dd}{\dd \tau_b}\left( \frac{\dd z_k}{\dd w_i}\big|_\tau\right)\;\frac{\dd \tau_b}{\dd \bar{w}_j}
-2 r_a\;\frac{\dd^2 \tau_a}{\dd w_i \dd \bar{w}_j}
\eea
We have shown above that $\frac{\dd}{\dd \tau_b}=\nabla_{\mathrm{J} \, V_b}$, therefore 
\bea
\frac{\dd}{\dd \tau_b}\frac{\dd \tilde{K}}{\dd z_k}= \big[i v_b^i \frac{\dd}{\dd z_i}-i \bar{v}_b^i \frac{\dd}{\dd \bar{z}_i}, \frac{\dd}{\dd z_k}\big]\,\tilde{K}+ \frac{\dd}{\dd z_j}\,\frac{\dd \tilde{K}}{\dd \tau_b}=
-i\frac{\dd v_b^i}{\dd z_k}\,\frac{\dd \tilde{K}}{\dd z_i}+2 \frac{\dd \mu_b}{\dd z_k}
\eea
Also note that
\bea
\frac{\dd}{\dd \tau_b}\left( \frac{\dd z_k}{\dd w_i}\big|_\tau\right)=\frac{\dd}{\dd w_i}\big|_\tau\left( \frac{\dd z_k}{\dd \tau_b}\right)=i \,\frac{\dd v^k_b}{\dd w_i}\big|_\tau=i \,\frac{\dd v^k_b}{\dd z_n}\,\frac{\dd z_n}{\dd w_i}\big|_\tau
\eea
Using the latter two equations in (\ref{form1}), we obtain
\bea\label{form2}
\frac{\dd^2 \tilde{K}}{\dd w_i \dd \bar{w}_j} =\frac{\dd^2 \tilde{K}}{\dd z_k \dd \bar{z}_m} \;\frac{\dd \bar{z}_m}{\dd \bar{w}_j}\big|_\tau\;\frac{\dd z_k}{\dd w_i}\big|_\tau+2 \frac{\dd \mu_b}{\dd z_k} \;\frac{\dd z_k}{\dd w_i}\big|_\tau\; \frac{\dd \tau_b}{\dd \bar{w}_j}
-2 r_a\;\frac{\dd^2 \tau_a}{\dd w_i \dd \bar{w}_j}
\eea
Differentiating the identity $\mu_b(z(w, \tau), \bar{z}(\bar{w}, \tau))=0$ w.r.t. $z_j$, we obtain
\bea
\frac{\dd \mu_b}{\dd z_k} \;\frac{\dd z_k}{\dd w_i}\big|_\tau=-\frac{\dd \mu_b}{\dd \tau_c}\;\frac{\dd \tau_c}{\dd w_i}
\eea
Plugging this in (\ref{form2}) we get
\bea
\frac{\dd^2 \tilde{K}}{\dd w_i \dd \bar{w}_j} =\frac{\dd^2 \tilde{K}}{\dd z_k \dd \bar{z}_m} \;\frac{\dd \bar{z}_m}{\dd \bar{w}_j}\big|_\tau\;\frac{\dd z_k}{\dd w_i}\big|_\tau-
2 \frac{\dd \mu_b}{\dd \tau_c}\;\frac{\dd \tau_c}{\dd w_i}\; \frac{\dd \tau_b}{\dd \bar{w}_j}
-2 r_a\;\frac{\dd^2 \tau_a}{\dd w_i \dd \bar{w}_j}
\eea
Comparing this with (\ref{derkah}), we need to show that
\bea\label{eq5}
\dd_k\bar{\dd}_s\tilde{K}\;\left( \frac{\dd z_k}{\dd w_i}\big|_{\bar{w}}\;\frac{\dd \bar{z}_s}{\dd \bar{w}_j}\big|_w-\frac{\dd z_k}{\dd \bar{w}_j}\big|_w\;\frac{\dd \bar{z}_s}{\dd w_i}\big|_{\bar{w}} \right) = \dd_k\bar{\dd}_s\tilde{K}\; \;\frac{\dd \bar{z}_s}{\dd \bar{w}_j}\big|_\tau\;\frac{\dd z_k}{\dd w_i}\big|_\tau-
2 \frac{\dd \mu_b}{\dd \tau_a}\;\frac{\dd \tau_a}{\dd w_i}\; \frac{\dd \tau_b}{\dd \bar{w}_j}
\eea
The derivatives with fixed $w, \bar{w}$ and fixed $\tau$ are related by
\bear \nonumber
&&\frac{\dd z_k}{\dd w_i}\big|_{\bar{w}}=\frac{\dd z_k}{\dd w_i}\big|_\tau+i\,v_a^k\frac{\dd \tau_a}{\dd w_i}, \quad
\frac{\dd \bar{z}_s}{\dd \bar{w}_j}\big|_w=\frac{\dd \bar{z}_s}{\dd \bar{w}_j}\big|_\tau-i\,\bar{v}^s_b \frac{\dd \tau_b}{\dd \bar{w}_j}\\ \nonumber
&&\frac{\dd z_k}{\dd \bar{w}_j}\big|_w=i\,v_a^k \frac{\dd \tau_a}{\dd \bar{w}_j}, \quad
\frac{\dd \bar{z}_s}{\dd w_i}\big|_{\bar{w}}= -i\,\bar{v}^s_b \frac{\dd \tau_b}{\dd w_i}
\eear
Plugging these expressions in the above formula and taking into account that
\bea
i\,v_a^k\frac{\dd \tau_a}{\dd w_i} \frac{\dd \bar{z}_s}{\dd \bar{w}_j}\big|_\tau \,\dd_k \bar{\dd}_s \tilde{K}=-i \bar{v}_b^s \frac{\dd \tau_b}{\dd \bar{w}_j} \frac{\dd z_k}{\dd w_i}\big|_\tau \,\dd_k \bar{\dd}_s \tilde{K}=-{1\over 2} \frac{\dd^2 \tilde{K}}{\dd \tau_a \dd \tau_b}\frac{\dd \tau_a}{\dd w_i}\frac{\dd \tau_b}{\dd \bar{w}_j}
\eea
we obtain the desired equality (\ref{eq5}). In the derivation we have also used the fact that
\bea
{1\over 2} \frac{\dd^2 \tilde{K}}{\dd \tau_a \dd \tau_b}=\frac{\dd \mu_b}{\dd \tau_a}=\frac{\dd \mu_a}{\dd \tau_a},
\eea
since $[{\dd \over \dd \tau_a}, {\dd \over \dd \tau_b}]\tilde{K}\sim f_{abc}\,\nabla_{V_c}\tilde{K}=0$.
Since $\tilde{K}(z(w, \tau),\bar{z}(\bar{w},\tau))$ is what we call $\pi^\ast (\tilde{K}|_{\mu^{-1}(0)})$, we have the desired result
\bea
\pi^{\ast}(\omega\big|_{\mu^{-1}(0)})=i\,\dd_w\;\bar{\dd}_{\bar{w}}\,\left(\pi^\ast (\tilde{K}|_{\mu^{-1}(0)})+2 \;\sum\limits_a\,r_a \,\tau_a \right)
\eea
so that the formula (\ref{Kquot}) holds:
\bea
\pi^\ast K=\pi^\ast (\tilde{K}|_{\mu^{-1}(0)})+2 \;\sum\limits_a\,r_a \,\tau_a
\eea

\section{The boundary correction \\in the Chern-Weil formulas}\label{ChernSimons}

Suppose $g_0$ is a metric, for which $\dd X$ is a totally geodesic submanifold of $X$, and $g$ is the metric of interest, in terms of which we wish to calculate the topological numbers. The corresponding connections will be called $\omega_0=\omega_0(g_0)$ and $\omega=\omega(g)$. We know how to calculate the topological numbers for the metric $g_0$ --- in that case, up to the so-called $\eta$-correction, they are given by the integrals (\ref{inteu}), (\ref{intsgn}). Let us call generally any of these integrands as $P(\omega_0)$. In that case
\bea
\textrm{Topological number}=\int\limits_X P(\omega_0)=\int\limits_X \big(P(\omega_0)-P(\omega)\big)+\int\limits_X P(\omega)
\eea
It turns out that $P(\omega_0)-P(\omega)=dU$ is exact, so that
\bea
\textrm{Topological number}=\int\limits_X P(\omega_0)=\int\limits_{\dd X} \,U+\int\limits_X P(\omega)
\eea
The second term is then the bulk contribution from the metric $g$ of interest, and the first term is the boundary contribution that we are after. Hence to evaluate the boundary contribution we need to find the form $U$ for various characteristic classes $P$ (more exactly, the Euler class and the Pontryagin class). Along the way for completeness we also prove that the difference $P(\omega)-P(\omega_0)$ is exact. Indeed, $P(\omega), P(\omega_0)$ both are closed and therefore locally exact, meaning that we can write $P(\omega)=dF(\omega)$, where $F(\omega)$ transforms nontrivially (inhomogeneously) under the gauge transformations of the connection $\omega$. The idea is that $F(\omega)-F(\omega_0)=G(\omega-\omega_0,R)$ is a function of the (gauge-covariant) curvature tensor $R$ and the \emph{difference} $\omega-\omega_0$, which is no longer a connection but rather a well-defined 1-form. To prove it we introduce a `line' in the space of connections
\bea
\omega_t=\omega+(1-t)(\omega_0-\omega)=\omega-\tau \zeta,\quad\textrm{where}\quad \tau=1-t,\;\zeta=\omega-\omega_0
\eea
 and write
\bea
P(\omega)-P(\omega_0)=\int\limits_{0}^1\,dt\,\frac{d P(\omega_t)}{dt}
\eea
In fact, for our purposes of computing $\mathrm{Eu}$ and $\mathrm{Sgn}$ we only need to consider two cases, namely
\bea
P=\tau_1={1\over 2}\tr(R^2)\quad \textrm{and} \quad P=\tau_2=\mathrm{Pf}(R)={1\over 8}\epsilon^{abcd}R_{ab}R_{cd},
\eea
where the curvature $R$ is determined from the connection via
\bea
R=d\omega+\omega \wedge \omega
\eea
Taking into account that it satisfies a Bianchi identity $dR+\omega \wedge R-R\wedge \omega=0$, we can calculate the derivative of $\tau_1(R_t)$ w.r.t. $t$:
\bea
\frac{d}{dt}\tau_1(R_t)=d\,\tr(R_t \zeta),
\eea
which leads to
\bea\label{tau1der}
-dU_1=\tau_1(R)-\tau_1(R_0)=d\,\big(\tr(R \zeta)-{1\over 2} \tr ( D\zeta \cdot \zeta) +{1\over 3} \tr(\zeta^3)\big),
\eea
where
\bea
D\zeta=d\zeta+\{\omega, \zeta\}
\eea
A similar calculation for $\tau_2(R_t)$ is slightly more complicated. To carry it out, first notice that $\langle A, B\rangle={1\over 4}\epsilon^{abcd}A_{ab}B_{cd}$ is an $SO(4)$-invariant pairing on the Lie algebra $\mathfrak{so}_4$ of skew-symmetric matrices. The fact that it is invariant means that $\langle g A g^\mathrm{T}, g B g^\mathrm{T}\rangle=\langle A, B\rangle$, which infinitesimally implies $\langle [C, A], B\rangle=-\langle A, [C,B]\rangle$. In what follows we will also employ an analogous formula for fermionic matrices (1-forms): $\langle \{ \xi, \chi\}, b\rangle=\langle \xi, [\chi, b]\rangle$, where $\xi, \chi$ are fermionic skew-symmetric matrices and $b$ is a bosonic skew-symmetric matrix\footnote{The `fermionic' formula can be derived from the `bosonic' one simply by assuming that $\xi=\epsilon_1 c$, $\chi = \epsilon_2 a$, where $\epsilon_{1,2}$ are Grassmann parameters and $a, c$ are bosonic matrices.}. Bearing this in mind, we may write\footnote{Here $\zeta^2={1\over 2} \{\zeta, \zeta\}$ is a shorthand for $\zeta\wedge \zeta$.}
\bear
&&\frac{d}{dt}\tau_2(R_t)=\langle D\zeta-2\tau\zeta^2, R-\tau \,D\zeta+\tau^2\zeta^2\rangle=\\ \nonumber
&&=\langle D\zeta, R\rangle-\tau (\langle D\zeta, D\zeta\rangle+2 \langle \zeta^2, R\rangle)+3\tau^2\langle D\zeta, \zeta^2\rangle-2 \tau^3 \langle \zeta^2, \zeta^2\rangle
\eear
Using the above formulas, we write
\bear\nonumber
&&\langle D\zeta, R \rangle =d\langle \zeta, R\rangle -\langle \zeta, d(\omega^2) \rangle+\langle \zeta, [\omega, R] \rangle =d\langle \zeta, R\rangle \\ \nonumber
&&\langle D\zeta, D\zeta\rangle+2 \langle \zeta^2, R\rangle=d\langle \zeta, D\zeta\rangle+\langle \zeta, [\omega, D\zeta]-[\omega, d\zeta]-[\zeta, d\omega]\rangle+2 \langle \zeta^2, R\rangle=d\langle \zeta, D\zeta\rangle
\\ \nonumber
&&\langle D\zeta, \zeta^2\rangle={1\over 3} d \langle \zeta, \zeta^2\rangle+\langle\omega, [\zeta, \zeta^2]\rangle={1\over 3} d \langle \zeta, \zeta^2\rangle\\ \nonumber
&&\langle \zeta^2, \zeta^2\rangle={1\over 2} \langle \zeta, [\zeta, \zeta^2]\rangle=0
\eear
Combining these results, we obtain:
\bea
\frac{d}{dt}\tau_2(R_t)=d\big( \langle \zeta, R\rangle- \tau \langle \zeta, D\zeta\rangle+\tau^2 \langle \zeta, \zeta^2\rangle\big)
\eea
Integrating over $t$, we get
\bea\label{tau2der}
-dU_2=\tau_2(R)-\tau_2(R_0)=d\big( \langle \zeta, R\rangle- {1\over 2} \langle \zeta, D\zeta\rangle+{1\over 3} \langle \zeta, \zeta^2\rangle \big)
\eea
The above expressions (\ref{tau1der}) and (\ref{tau2der}) may be simplified in the case of interest, where the boundary is given by the equation $r=r_{\textrm{boundary}}$ (fixed radius), or equivalently by the condition $E_0=0$, where $E_0$ is a component of the vierbein (see Appendix \ref{metric} below). In this case the only nonzero components of $\zeta$ at the boundary are $\zeta_{0i}$ and $\zeta_{i0}=-\zeta_{0i}$. Moreover, $\zeta_{0i}|_{\textrm{boundary}}=\omega_{0i}|_{\textrm{boundary}}$. Since $R_{0i}=d\omega_{0i}+\omega_{0k}\wedge \omega_{ki}$, we can also express $d\omega_{0i}$ through $\omega$ and $R$. At the end we obtain the following result
\bear
&&U_1=-\frac{1}{2}\,\tr(\zeta\, R)\\
&&U_2=-\langle \zeta, R\rangle+{2\over 3} \langle \zeta, \zeta^2\rangle,
\eear
which gives rise to formulas (\ref{delta1}), (\ref{delta2}).

\section{The $U(2)$-invariant metrics in explicit form}\label{metric}

In this Appendix we wish to write out the line element of a metric on $X$ originating from a K\"ahler potential that is a function of $x=r^2=|z_1|^2+|z_2|^2$: $K=K(x)$. In fact, below it will be convenient to use the function $Q(x)=xK'$ in place of $K$. We write the complex coordinates $(z_1, z_2)$ as
\bea
z_1 = r u_1,\quad z_2= r u_2,\quad\textrm{where}\quad |u_1|^2+|u_2|^2=1
\eea
$(u_1, u_2)$ parametrize an $S^3=SU(2)$, so it is convenient to introduce an $SU(2)$ group element
\bea
g=\left(\scalemath{1}{
\begin{array}{cc}
u_1&-\bar{u}_2\\u_2&\bar{u}_1
\end{array}}\right)
\eea
Next, we introduce a triad of left-invariant 1-forms:
\bea
e_i:=i\,\tr(\sigma_i g^\dagger dg),\;\;i=1, 2, 3.
\eea
Here $\sigma_i$ are the Pauli matrices. For convenience we write out this triad explicitly:
\bear
&&e_1=i(u_1 du_2-u_2 du_1+\bar{u}_2 d\bar{u}_1-\bar{u}_1 d\bar{u}_2)\\ \nonumber
&&e_2=u_1 du_2-u_2 du_1+\bar{u}_1 d\bar{u}_2-\bar{u}_2 d\bar{u}_1\\ \nonumber
&&e_3=2i(\bar{u}_1 du_1+\bar{u}_2 du_2)
\eear
The metric then has the following form:
\bea\label{lineel}
4\,ds^2= \frac{Q'}{x}\,dx^2+Q(e_1^2+e_2^2)+xQ' e_3^2
\eea
We can write the line element as $ds^2=\sum\limits_{i=0}^3\,E_i^2$, where the vierbein $E_i$ is given by
\bea\label{vierbein}
E_0={1\over 2}\sqrt{\frac{Q'}{x}}\,dx,\quad E_1={1\over 2} \sqrt{Q}\,e_1,\quad E_2={1\over 2} \sqrt{Q} e_2,\quad E_3={1\over 2} \sqrt{x Q'}\,e_3
\eea
The vierbein (\ref{vierbein}) satisfies Cartan's equations
\bea
dE_i+\sum\limits_{j=0}^3\,\omega_{ij} \wedge E_j=0,
\eea
where the connection $\omega$ represented by the skew-symmetric matrix $\omega_{ij}=-\omega_{ji}$ is as follows:
\bear
&&\omega_{01}=-\frac{\sqrt{x Q'}}{Q} E_1,\quad \omega_{02}=-\frac{\sqrt{x Q'}}{Q} E_2,\quad \omega_{03}=-\frac{2(\sqrt{x Q'})'}{Q'}\,E_3\\ \nonumber
&&\omega_{12}=-\big( \frac{2}{\sqrt{x Q'}}-\frac{\sqrt{x Q'}}{Q}\big)\,E_3,\quad \omega_{13}= \frac{\sqrt{x Q'}}{Q}\,E_2,\quad\omega_{23}= \frac{\sqrt{x Q'}}{Q}\,E_1
\eear
\appendix
\renewcommand\refname{\begin{center}
\vspace{-0.5cm}
\line(1,0){150}\\
{\normalfont\scshape \large References}\\
\vspace{-0.2cm}
\line(1,0){150}
\end{center}}
\bibliography{refskah}
\bibliographystyle{ieeetr}

\end{document}